%% file: main.tex
\documentclass[sigconf,nonacm]{acmart}

\usepackage{cancel}     
\usepackage{xurl}       
\usepackage{subfigure}  
\usepackage{siunitx}    
\usepackage{enumitem}   
\usepackage{xcolor}     
\usepackage{xspace}     

\usepackage{algorithm}
\usepackage[noend]{algpseudocode}
\usepackage{etoolbox}
\let\ReturnInline\Return
\renewcommand{\Return}{\State\ReturnInline}
\algrenewcommand\algorithmicrequire{$\rhd$}
\algrenewcommand\algorithmicensure{$\square$}

\input{macros}

\AtBeginDocument{%
  \providecommand\BibTeX{{%
    \normalfont B\kern-0.5em{\scshape i\kern-0.25em b}\kern-0.8em\TeX}}}

\setcopyright{acmcopyright}
\copyrightyear{2024}
\acmYear{2024}
\acmDOI{XXXXXXX.XXXXXXX}

\begin{document}

\title{Lock-Free Computation of PageRank in Dynamic Graphs}


\author{Subhajit Sahu}
\email{subhajit.sahu@research.iiit.ac.in}
\orcid{0000-0001-5140-6578}
\affiliation{%
  \institution{IIIT Hyderabad}
  \streetaddress{Professor CR Rao Rd, Gachibowli}
  \city{Hyderabad}
  \state{Telangana}
  \country{India}
  \postcode{500032}
}


\settopmatter{printfolios=true}

\begin{abstract}
PageRank is a\ignore{popular} metric that assigns importance to the vertices of a graph based on its neighbors and their scores. Recently, there has been increasing interest in computing PageRank on dynamic graphs, where the graph structure evolves due to edge insertions and deletions. However, traditional barrier-based approaches for updating PageRanks encounter significant wait times on certain graph structures, leading to high overall runtimes. Additionally, the growing trend of multicore architectures with increased core counts has raised concerns about random thread delays and failures. In this study, we propose a lock-free algorithm for updating PageRank scores on dynamic graphs. First, we introduce our Dynamic Frontier (DF) approach, which identifies and processes vertices likely to change PageRanks with minimal overhead. Subsequently, we integrate DF with our lock-free and fault-tolerant PageRank (\FroBarf{}), incorporating a helping mechanism among threads between its two phases. Experimental results demonstrate that \FroBarf{} not only eliminates waiting times at iteration barriers but also withstands random thread delays and crashes. On average, it is $4.6\times$ faster than lock-free Naive-dynamic PageRank (\NaiBarf{}).
\ignore{PageRank is a popular centrality metric that assigns importance to the vertices of a graph based on its neighbors and their score. In recent years, computing PageRank on dynamic graphs is gaining research attention as the graphs of interest evolve due to the insertion and deletion of edges. In addition, the architectural trend of increasing the core count on multicore architectures has introduced concerns on random delays and failures of threads.}
\ignore{This paper addresses the design of a fault-tolerant PageRank algorithm in the batch dynamic setting. First, our \textit{Static Barrier-free} PageRank is presented, which tolerates random thread delays/crashes with a simpler design when compared to existing work, and is at least $14\%$ faster. This is adapted to two well-known dynamic PageRank algorithms.}
\ignore{Next, our \textit{Dynamic Frontier} approach is discussed. Given a batch update of edge deletion and insertions, it progressively identifies affected vertices that are likely to change their ranks with minimal overhead. We integrate this approach with our improved \textit{Barrier-free} PageRank to arrive at a fault-tolerant and high-performance implementation. Experimental results on $12$ real-world graphs indicate that our \textit{Dynamic Frontier Barrier-free} PageRank is on average $4.6\times$ faster than existing dynamic approaches. Simulated fault injection studies indicate that our proposed approach achieves good performance in the presence of random thread delays and also tolerates random thread crashes.}
\end{abstract}



\maketitle

\section{Introduction}
\label{sec:introduction}
\input{01-introduction}

\section{Related work}
\label{sec:related}
\input{05-related-work}

\section{Preliminaries}
\label{sec:preliminaries}
\input{02-preliminaries}

\section{Approach}
\label{sec:approach}
\input{03-approach}

\section{Evaluation}
\label{sec:evaluation}
\input{04-evaluation}

\section{Conclusion}
\label{sec:conclusion}
\input{06-conclusion}

\begin{acks}
I would like to thank Prof. Kishore Kothapalli, Prof. Hemalatha Eedi, and Prof. Sathya Peri for their support.
\end{acks}

\bibliographystyle{ACM-Reference-Format}
\bibliography{main}

\clearpage
\appendix
\section{Appendix}
\input{aa-appendix}
\end{document}

%% file: macros.tex
\newcommand{\punt}[1]{}
\newcommand{\cmnt}[1]{}
\newcommand{\ignore}[1]{}

\algnewcommand{\IIf}[1]{\State\algorithmicif\ #1\ \algorithmicthen}
\algnewcommand{\EndIIf}{\unskip\ \algorithmicend\ \algorithmicif}

\definecolor{darkblue}{rgb}{0.0, 0.0, 0.55}

\newcounter{history}

\algdef{SE}[DOWHILE]{Do}{doWhile}{\algorithmicdo}[1]{\algorithmicwhile\ #1}%
\newcommand{\lf}{lock-free\xspace}

\newcommand{\lfdm}{lock-freedom\xspace}

\newcommand{\StaWbar}{$\text{Static}_{\text{BB}}$}
\newcommand{\StaBarf}{$\text{Static}_{\text{LF}}$}
\newcommand{\NaiWbar}{$\text{ND}_{\text{BB}}$}
\newcommand{\NaiBarf}{$\text{ND}_{\text{LF}}$}
\newcommand{\TraWbar}{$\text{DT}_{\text{BB}}$}
\newcommand{\TraBarf}{$\text{DT}_{\text{LF}}$}
\newcommand{\FroWbar}{$\text{DF}_{\text{BB}}$}
\newcommand{\FroBarf}{$\text{DF}_{\text{LF}}$}

\makeatletter
\newcommand*{\algrule}[1][\algorithmicindent]{%
  \makebox[#1][l]{%
    \hspace*{.2em}
    \vrule height .75\baselineskip depth .25\baselineskip
  }
}

\newcount\ALG@printindent@tempcnta
\def\ALG@printindent{%
    \ifnum \theALG@nested>0
    \ifx\ALG@text\ALG@x@notext
    \else
    \unskip
    \ALG@printindent@tempcnta=1
    \loop
    \algrule[\csname ALG@ind@\the\ALG@printindent@tempcnta\endcsname]%
    \advance \ALG@printindent@tempcnta 1
    \ifnum \ALG@printindent@tempcnta<\numexpr\theALG@nested+1\relax
    \repeat
    \fi
    \fi
}
\patchcmd{\ALG@doentity}{\noindent\hskip\ALG@tlm}{\ALG@printindent}{}{\errmessage{failed to patch}}
\patchcmd{\ALG@doentity}{\item[]\nointerlineskip}{}{}{} 
\makeatother

%% file: 01-introduction.tex
The increasing availability of vast amounts of data represented as graphs has led to significant interest in algorithms for addressing graph-related problems. One prominent algorithm in this domain is PageRank, which assigns a score indicating the importance of vertices within a graph. PageRank is particularly well-known for its application in ranking web pages within search engines such as Google \cite{rank-page99}, but its utility extends to a number of other fields. These include identifying misinformation, predicting traffic flow \cite{traffic-kim15}, urban planning \cite{urban-zhang18}, in recommendation systems \cite{recommend-chaudhari17}, protein target identification \cite{banky2013equal}, and in neuroscience \cite{rank-gleich15}.

Due to the algorithm's popularity and the challenges posed by the scale of modern data, there is substantial research dedicated to designing and implementing parallel algorithms for computing PageRank. These efforts leverage a variety of hardware platforms, including multicore CPUs \cite{rank-garg16}, GPUs \cite{rapids}, FPGAs \cite{rank-guoqiang20}, SpMV ASICs \cite{rank-sadi18}, and hybrid systems combining CPUs with GPUs or FPGAs \cite{rank-giri20, rank-li21}. Distributed systems also play a crucial role in handling large-scale data \cite{rank-sarma13}. The techniques employed in these implementations often involve power iterations, random walks, and Markov chains to efficiently compute the PageRank scores.

Despite advancements in PageRank\ignore{algorithms}, three critical challenges persist. The first concerns dynamic graphs. Here, the underlying graph evolves with time through insertion and deletion of vertices and edges. However, it is generally impractical to resort to a full recomputation of PageRank on small changes to the underlying graph. Efficient algorithms for dynamic graphs should only recalculate the PageRanks of potentially affected vertices ---  whose ranks are likely to change. Current approaches \cite{rank-zhang17, rank-desikan05, kim2015incremental, rank-giri20, rank-sahu22} either reprocess all vertices or those reachable from the updated graph region, thereby processing numerous vertices even for minor updates.

The second challenge concerns prolonged wait times in parallel barrier-based algorithms for computing the PageRank \cite{rank-garg16}. These algorithms employ edge-balanced or vertex chunking strategies for load balancing. In \textit{edge-balanced} algorithms, the edges are evenly distributed between threads while in \textit{vertex chunking} algorithms vertices are grouped into chunks of fixed size and dynamically distributed between threads. In both cases, threads execute synchronous iterations and wait at a barrier for straggler threads at the end of each iteration. The vertex chunking strategy is favored due to its avoidance of pre-processing and atomic operations \cite{heidari2018scalable}. However, specific graph structures can lead to significant wait times with vertex chunking, despite efforts to optimize work scheduling. It can be seen from Figure \ref{fig:adjust-chunk} that reducing chunk size reduces waiting time. But it also increases overall computation time due to scheduling overheads. Figure \ref{fig:adjust-chunk} specifically shows that the thread wait time at barriers can make up to $73\%$ of total execution time for computing vertex PageRanks. This trade-off between waiting time and computation time highlights the\ignore{inherent} inefficiencies in barrier-based PageRank computation algorithms. 

\input{src/fig-adjust-chunk}

The third issue pertains to architectural concerns of modern hardware. While modern hardware offers significant working memory size and performance, architectural advances in CPU complexity, ever increasing core counts, and limits of CMOS scaling introduce challenges in maintaining reliability, especially in high-performance computer systems. Engineers must carefully balance fault resilience and cost efficiency in their designs \cite{fault-bridges12}. Notably, the work of Hochschild et al. \cite{fault-hochschild21} indicates that mercurial cores --- those exhibiting intermittent faults --- are observed at a rate of a few cores per several thousand machines, with this rate gradually increasing. Typically, only one core in a multi-core processor fails, and such faulty cores fail intermittently and progressively worsen over time, an issue prevalent across the industry, varying by CPU product and influenced by factors like frequency, voltage, and temperature. These core faults, undetected during manufacturing tests, can lead to hard faults, such as thread crashes and program interruptions; or soft faults, such as random thread delays\ignore{, which can manifest unpredictably and often worsen with machine aging \cite{fault-hochschild21}}. Additionally, arbitrary delays can stem from contention, false sharing, memory access delays induced by page faults or cache coherence protocols, or overheating CPU cores, leading to a drop in performance. These issues also cause slower straggler threads, resulting in a further drop in performance for barrier-based algorithms.\ignore{For PageRank computation of massive graphs, it is important to tolerate faults - faults become increasingly likely at this stage.} Going forward, algorithms are expected to tolerate such faults in exchange for obtaining higher performance and energy efficiency.

A multi-threaded lock-free algorithm \cite{Herlihy+:nature:opodis:2011, Herlihy+:Art:Book:2020} guarantees non-blocking progress for all threads, preventing a slow or crashed thread from hindering others and ensuring algorithm termination. This property, known as lock-freedom, is crucial for dynamic graph algorithms, and is achieved without locks or barriers, which typically introduce blocking. Eedi et al. \cite{rank-eedi22} recently developed a lock-free algorithm for computing PageRank. However, their method only works for static graphs\ignore{ and do not work on dynamic graphs}.\ignore{Mitigating such behavior via dynamic scheduling or work stealing falls short in critical situations or requires additional bookkeeping.}\ignore{Banerjee et al. \cite{rank-sahu22} studied parallel dynamic algorithms for computing PageRank using techniques developed by Garg et al. \cite{rank-garg16}. The algorithms from \cite{rank-sahu22} rely on dynamic scheduling of threads but fail to finish the computation even when a single thread crashes.}

\input{src/fig-about-delay}
\input{src/fig-about-crash}

Figures \ref{fig:about-delay} and \ref{fig:about-crash} show the behavior of barrier-based and lock-free PageRank in the presence of faults. Here, $th_i$ represents a\ignore{running} thread, $Ci$ represents the time taken for processing each chunk of vertices, and a solid/dotted vertical line represents the end of an iteration. In barrier-based PageRank with random thread delays, threads unaffected by a random thread delay must wait for the delayed threads to reach the end of the barrier. See for example Figure \ref{fig:about-delay-01}, where $th_2$ which does not experience a thread delay must wait for $th_1$\ignore{which does experience a thread delay}. However, with lock-free PageRank, the threads are able to make progress independently of each other, and are not required to wait for each other at the end of the barrier. This is shown in Figure \ref{fig:about-delay-02}, where $th_2$ is able to make progress independently of $th_1$. In a similar manner, in the presence of random thread crashes, a non-crashed thread is unable to make progress as it continues to wait for the crashed thread to reach the barrier. This is shown in Figure \ref{fig:about-crash-01}, where $th_2$ must wait for $th_1$ to reach the barrier. However, with lock-free PageRank, the threads are able to make progress independently, even in the presence of random thread crashes. This is shown in Figure \ref{fig:about-crash-02}, where $th_2$ is able to make progress even when $th_1$ crashes.

\ignore{From the above discussion, we note a need to design parallel algorithms for obtaining the PageRank values of nodes in a dynamic graph. In particular, algorithms and implementations for dynamic parallel graph algorithms that are lock-free and hence can tolerate various thread failures, such as thread crashes and delays, are essential. In this paper, we show that designing such implementations for iterative algorithms, such as the PageRank algorithm, is efficient for dynamic graphs.}

\subsection{Our Contributions}

This technical report addresses the design of a fault-tolerant PageRank algorithm in the batch dynamic setting, where multiple edge updates are processed simultaneously. We summarize our contributions in the following:

\begin{itemize}
  \item In Section \ref{sec:frontier}, we propose the Dynamic Frontier (DF) approach for updating PageRanks on dynamic graphs. It progressively identifies and processes vertices whose PageRanks are likely to change due to a batch update.
  \item In Sections \ref{sec:frontier-withbarrier} and \ref{sec:frontier-barrierfree}, we present a barrier-based (\textbf{\FroWbar{}}) and a lock-free implementation (\textbf{\FroBarf{}}) of the DF approach. \FroBarf{} tolerates random thread delays and crashes\ignore{gracefully}.
  \item In Section \ref{sec:fault-none}, experimental results on two real-world dynamic graphs and twelve real-world large graphs with generated random updates demonstrate that \FroBarf{} is on average $4.6\times$ faster than a Naive-dynamic approach (\textbf{\NaiBarf{}}). 
  \item In Sections \ref{sec:fault-sleep} and \ref{sec:fault-crash}, we demonstrate the fault tolerance of \FroBarf{} by running it under simulated thread delays and crashes. Our results indicate that \FroBarf{} converges\ignore{achieves convergence}, with a gradual performance degradation as the\ignore{delays/number of} faults increase. 
\end{itemize}

The codebase is available online.\footnote{\url{https://github.com/puzzlef/pagerank-barrierfree-openmp-dynamic}}

%% file: src/fig-adjust-chunk.tex
\begin{figure}[!hbt]
  \centering
  \subfigure{
    \label{fig:adjust-chunk--all}
    \includegraphics[width=0.99\linewidth]{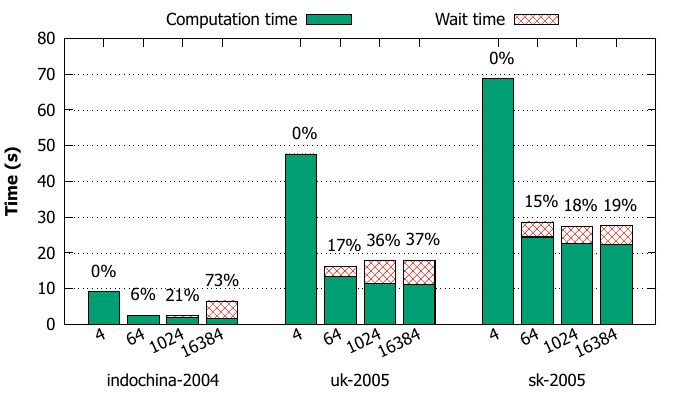}
  } \\[-2ex]
  \caption{Computation time and waiting time of barrier-based Static PageRank algorithm on three large graphs, using dynamic work scheduling of vertex chunks with varying sizes from 4 to 16,384 in multiples of 16. Labels on the bars indicate the percentage wait time.}
  \label{fig:adjust-chunk}
\end{figure}

%% file: src/fig-about-delay.tex
\begin{figure}[hbtp]
  \centering
  \begin{minipage}{0.75\linewidth}
  \subfigure[With random thread delays (barrier-based)]{
    \label{fig:about-delay-01}
    \includegraphics[width=0.99\linewidth]{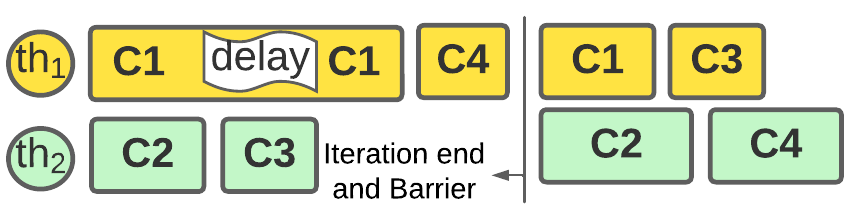}
  }
  \subfigure[With random thread delays (lock-free)]{
    \label{fig:about-delay-02}
    \includegraphics[width=0.99\linewidth]{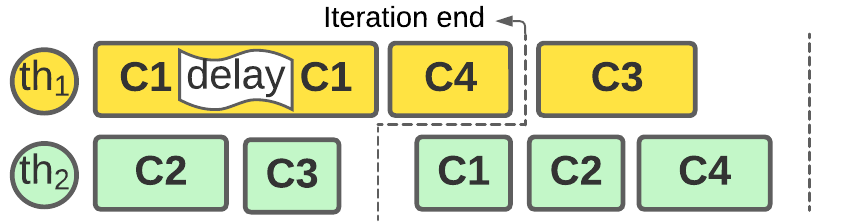}
  }
  \end{minipage} \\[-1ex]
  \caption{An example showing the behavior of barrier-based and lock-free PageRank on a graph with four vertex chunks, C1 through C4, in the presence of random thread delays.}
  \label{fig:about-delay}
\end{figure}

%% file: src/fig-about-crash.tex
\begin{figure}[hbtp]
  \centering
  \begin{minipage}{0.75\linewidth}
  \subfigure[With random thread crashes (barrier-based)]{
    \label{fig:about-crash-01}
    \includegraphics[width=0.99\linewidth]{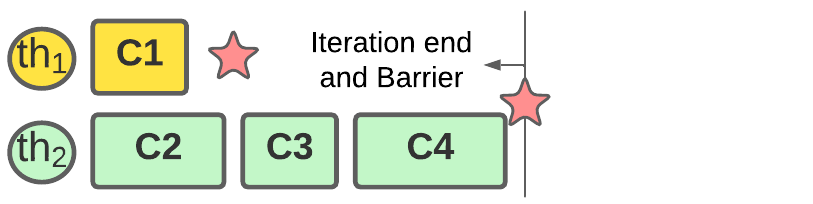}
  }
  \subfigure[With random thread crashes (lock-free)]{
    \label{fig:about-crash-02}
    \includegraphics[width=0.99\linewidth]{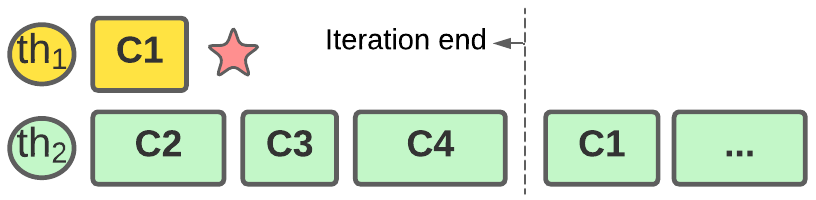}
  }
  \end{minipage} \\[-1ex]
  \caption{An example illustrating barrier-based and lock-free PageRank on a graph with four vertex chunks, C1 to C4, amidst random thread crash-stops (indicated with a star).}
  \label{fig:about-crash}
\end{figure}

%% file: 05-related-work.tex
Parallel algorithms for problems on dynamic graphs have been attracting a lot of research attention in recent years. Examples include the work of Regunta et al. \cite{cent-regunta21}, Sahu et al. \cite{rank-sahu22}, Haryan et al. \cite{cc-haryan22}, and Khanda et al. \cite{khanda22parallel}. Some of the early work on dynamic graph algorithms in the sequential setting include the seminal sparsification method of Eppstein et al. \cite{graph-eppstein97} and the bounded incremental computation idea of Ramalingam \cite{incr-ramalingam96}.\ignore{The latter advocates measuring the work done as part of the update in proportion to the effect the update has on the computation.}

PageRank is a fundamental algorithm used to measure relationships among vertices and subsets of nodes in various applications. It has been implemented on multicore CPUs \cite{rank-garg16}, GPUs \cite{rapids}, FPGAs \cite{rank-guoqiang20}, SpMV ASICs \cite{rank-sadi18}, CPU-GPU hybrids \cite{rank-giri20}, CPU-FPGA hybrids \cite{rank-li21}, and distributed systems \cite{rank-sarma13}.

Several approaches have been proposed to incrementally compute approximate PageRank values in dynamic/evolving graphs. Chien et al. \cite{rank-chien01} identify a small region near updated vertices in the graph, condense the rest of the graph into a single vertex, compute PageRanks for this condensed graph, and then map them back to the original graph. Chen et al. \cite{chen2004local} trace backward from the target node along reverse hyperlinks, to estimate its PageRank score. Bahmani et al. \cite{bahmani2010fast}, Zhan et al. \cite{zhan2019fast}, and Pashikanti et al. \cite{rank-pashikanti22} present Monte Carlo-based algorithms for\ignore{approximate} PageRank tracking on dynamic networks. Zhang \cite{rank-zhang17} utilizes a simple incremental PageRank computation method, which we refer to as the \textit{Naive-dynamic (ND)} approach, on hybrid CPU and GPU platforms\ignore{that incorporates the Update-Gather-Apply-Scatter (UGAS) computation model}.

A prevalent approach for updating PageRank \cite{rank-desikan05, kim2015incremental, rank-giri20, rank-sahu22} involves identifying the affected region of the graph after modifications, like edge insertions or deletions, using Breadth-First Search (BFS) or Depth-First Search (DFS) traversal from vertices connected to the modified edges. PageRanks are then computed exclusively for this region. We call this the \textit{Dynamic Traversal (DT)} approach. Kim and Choi \cite{kim2015incremental} apply this\ignore{approach} with an asynchronous PageRank implementation, while Giri et al. \cite{rank-giri20} employ it with collaborative executions on multi-core CPUs and massively parallel GPUs. Sahu et al. \cite{rank-sahu22} utilize this on a Strongly Connected Component (SCC)-based graph decomposition to confine computation to reachable SCCs from updated vertices, on multi-core CPUs and GPUs.\ignore{Some of the existing papers \cite{yang2020graphabcd, sha2017technical} address challenges like low convergence, improper load balancing in barrier-based computations.}

However, the above proposed algorithms are neither lock-free, nor fault tolerant. Herlihy and Shavit \cite{Herlihy+:nature:opodis:2011} emphasize the need for lock-free algorithms that are resilient to arbitrary thread slowdown and failures in parallel computing environments. Cong and Bader \cite{lock-cong05} propose a parallel lock-free Shiloach-Vishkin algorithm for finding connected components and parallel block-free (spinlock-based) Boruvka algorithm for finding a minimum spanning tree. They show that lock-free implementations can handle large graphs and have superior performance. Checkpointing is another strategy for handling failures in iterative graph algorithms. But it has high overhead costs \cite{graph-xu16, graph-xu18, graph-yan16}.\ignore{Optimized checkpointing strategies have been proposed, such as selective checkpointing \cite{graph-wang17}, lightweight checkpoint \cite{graph-yan16}, non-blocking checkpoint \cite{graph-xu16}, dynamically adjusted checkpoint intervals \cite{graph-wang17}, and injected checkpoints \cite{graph-xu16}.} In contrast, checkpoint-free strategies exploit algorithmic properties to achieve failure recovery \cite{graph-xu18}.

%% file: 02-preliminaries.tex
\subsection{PageRank algorithm}
\label{sec:pagerank}

The PageRank, $R[v]$, of a vertex $v \in V$ in the graph $G(V, E)$, represents its \textit{importance} and is based on the number of incoming links and their significance. Equation \ref{eq:pr} shows how to calculate the PageRank of a vertex $v$ in the graph $G$, with $V$ as the set of vertices ($n = |V|$), $E$ as the set of edges ($m = |E|$), $G.in(v)$ as the incoming neighbors of vertex $v$, $G.out(v)$ as the outgoing neighbors of vertex $v$, and $\alpha$ as the damping factor. Each vertex starts with an initial PageRank of $1/n$. The \textit{power-iteration} method updates these values iteratively until the change is rank values is within a specified tolerance $\tau$ value (indicating that convergence has been achieved).

\ignore{The PageRank vector $R$ can be written in matrix form as $R_0 \mathbf{M}^\infty$, where $R_0$ is the initial PageRank vector (where the PageRank of each vertex is $1/|V|$), $\mathbf{M} = \alpha \mathbf{P} + \frac{1 - \alpha}{|V|} \mathbf{J}$ is the transition matrix, $\mathbf{J}$ is an all ones matrix, and $\mathbf{P}$ is the normalized adjacency matrix of the graph (where $\mathbf{P}_{ij} = 1/d_i$ if there is an edge between vertices $i$ and $j$, and $d_i$ is the out-degree of vertex $i$). PageRank converges on a transition matrix $\mathbf{M}$ that is stochastic, irreducible and primitive. A transition matrix $\mathbf{M}$ is stochastic if none of the rows are all zeros, i.e., there are no dead ends. This can be solved by adding self-loops to all vertices. Further, $\mathbf{M}$ is irreducible if every node is reachable from every other node (when interpreted as a graph), and it is primitive if $\mathbf{M}^k > 0$ for some $k$. The use of a damping factor makes PageRank irreducible, primitive, and also avoids the undesirable effect of spider-traps, which\ignore{point to each other and} would otherwise capture all PageRank \cite{rank-langville06}.}

The \textit{random surfer model} is a conceptual framework for the PageRank algorithm, where a random surfer moves through the web by following the links on each page. The damping factor, $\alpha$, is the probability that the random surfer will continue to the next page along one of the links, instead of jumping to a random page on the web, and has a default value of $0.85$. The PageRank of each page can be seen as the long-term probability that the random surfer will visit that page, given that he starts on a random page and follows links according to the damping factor. The PageRank values can be calculated by finding the eigenvector of a transition matrix that represents the probabilities of moving from one page to another\ignore{in the Markov Chain}.

\begin{equation}
\label{eq:pr}
    R[v] = \alpha \times \sum_{u \in G.in(v)} \frac{R[u]}{|G.out(u)|} + \frac{1 - \alpha}{n}
\end{equation}

\subsection{Lock-free Algorithms}

In the context of parallel programming, a barrier is a program construct where each thread must wait until \textit{all other} threads progress in their execution up to a certain pre-designed position of the code. While such a construct simplifies the design of parallel algorithms, and makes the algorithms easier to reason about, threads that reach the barrier will necessarily wait at the barrier for slow threads to make progress up to the barrier. This introduces inefficiency in the program execution. Further, if some thread crashes --- but in the absence of undefined/byzantine behavior, i.e., simply causing termination of the thread without affecting memory contents of the process in an unexpected manner (crash-stop model) --- all the other threads will be be in a deadlocked state, unable to make progress. The other synchronization construct commonly used are locks which also have similar disadvantages.

As pointed out earlier, with modern architectures, engineers must design their algorithms to be fault-tolerant in order to reap the benefits of increased memory and performance, that are necessary when working with large datasets. It is therefore important that parallel algorithms are designed such that progress is always guaranteed, even in the presence of random thread delays or crashes --- while avoiding the use of barriers. Such algorithms are known as lock-free algorithms, and are gaining research attention \cite{lock-cong05,rank-eedi22}.

\subsection{Parallel Algorithms for Static Graphs}
\label{sec:about-static}

\subsubsection{Barrier-based PageRank \cite{rank-garg16, rank-sahu22, rank-giri20}}
\label{sec:withbarrier}

This is the standard parallel implementation of PageRank. Here, the main thread spawns multiple threads at the start of each iteration, where each thread computes PageRanks for a subset of graph vertices. Threads synchronize at a barrier after each iteration, until convergence.\ignore{This process repeats until the ranks converge, i.e., the change in ranks lies within a specified tolerance $\tau$.} The computation of ranks of vertices is synchronous in each iteration, by maintaining the updated ranks in a separate vector, similar to the \textit{Jacobi} method. This helps avoid interference between threads and variability in the final result.\ignore{Dynamic scheduling in OpenMP allows threads to select work from a global pool \cite{openmp-chapman07} and aim for work balance among threads.} We denote this as \textbf{\StaWbar{}}. The pseudocode of \StaWbar{} is given in Algorithm \ref{alg:with-barrier-static}.

However, \StaWbar{} is vulnerable to inefficiencies as shown in Figure \ref{fig:adjust-chunk} which shows that the performance can be slowed by the presence of high-degree vertex chunks that hinder load-balancing. Further, as shown in Figure \ref{fig:about-delay}, a slow thread or a thread experiencing arbitrary delays can cause all the other threads to slow down at the barrier. Our experiments reveal that when the duration of a random thread delay is large relative to the iteration time, the performance of \StaWbar{} is impacted. Additionally, even if a single thread crashes, the iteration barrier prevents other threads from advancing. If the thread delay is really large, it can be seen similar to a crash failure.

\subsubsection{Lock-free PageRank \cite{rank-eedi22}}
\label{sec:barrierfree-static}

Eedi et al. \cite{rank-eedi22} recently proposed a lock-free variant of the PageRank algorithm for static graphs which does not use barriers. In this approach, each thread processes a subset of vertices independently without waiting for other threads to complete an iteration. If some threads straggle or crash, the remaining threads help each other, ensuring overall progress and enabling the algorithm to reach convergence. The PageRank values are computed asynchronously --- in-place using a single vector, similar to the \textit{Gauss-Seidel} method --- allowing updates made by a thread to be visible to other threads within the same iteration.

However, the original algorithm by Eedi et al. uses static scheduling, where each thread processes a fixed set of vertices. This approach can lead to workload imbalances and slower convergence, as different regions of the graph progress unevenly --- in addition to requiring additional machinery to be fault-tolerant to random thread delays and crash-stops. To address these issues, we implement a modified lock-free PageRank with dynamic scheduling, referred to as \textbf{\StaBarf{}}. Our implementation utilizes a top-level parallel block, OpenMP's dynamic loop scheduling with a chunk size of $2048$ with the \texttt{nowait} directive, and a single shared PageRank vector. Dynamic loop scheduling causes each running thread to obtain a chunk of vertices to be processed from a global work pool, resulting in the running threads having a balanced workload. This implementation does not use barriers, and is lock-free since threads always make overall progress in PageRank computation of all the vertices. To check for convergence, we use an additional flag vector $R_C$. The psuedocode of \StaBarf{} is given in Algorithm \ref{alg:barrier-free-static}. In the absence of faults, our experiments show this implementation is $14\%$ faster than Eedi et al.'s \textit{No-Sync} version of barrier-free PageRank.

\subsection{Dynamic Graphs}
\label{sec:about-dynamic}

A dynamic graph can be viewed as a sequence of graphs, where $G^t(V^t, E^t)$ denotes the graph at time step $t$. The changes between graphs $G^{t-1}(V^{t-1}, E^{t-1})$ and $G^t(V^t, E^t)$ at consecutive time steps $t-1$ and $t$ can be denoted as a batch update $\Delta^t$ at time step $t$ which consists of a set of edge deletions $\Delta^{t-} = \{(u, v)\ |\ u, v \in V\} = E^{t-1} \setminus E^t$ and a set of edge insertions $\Delta^{t+} = \{(u, v)\ |\ u, v \in V\} = E^t \setminus E^{t-1}$. For simplicity, in this report, we assume that no vertices are added or removed from the graph. Ranks of vertices need to appropriately scaled when vertices deletions and insertions and involved.

\paragraph{Interleaving of graph update and computation:}

We assume that graph updates arrive in batches, with graph updates and algorithm execution being interleaved. If graph updates are needed during algorithm execution, a read-only snapshot \cite{graph-dhulipala19} of the graph is required for subsequent algorithm execution.\ignore{Changes to the graph arrive in a batched manner, with updating of the graph and execution of the desired algorithm being interleaved (i.e., there is only one writer upon the graph at a given point of time). In case it is desirable to update the graph while an algorithm is still running, a snapshot of the graph needs to be obtained, upon which the desired algorithm may be executed. See for example Aspen graph processing framework which significantly minimizes the cost of obtaining a read-only snapshot of the graph \cite{graph-dhulipala19}.}\ignore{This is an experimental paper of PageRank algorithm, we do not focus on graph update. It would be possible to perform PageRank computation and graph update in parallel. However, we choose to focus on updating PageRanks on dynamic batch updates on a snapshot of the graph.}\ignore{Computation is generally performed on a graph snapshot that does not change with time.}

\subsection{Algorithms for Updating PageRank on Dynamic Graphs}

\subsubsection{Naive-dynamic (ND) approach \cite{rank-langville06, rank-zhang17}}
\label{sec:about-naive}

This is a basic approach of updating PageRanks of vertices in dynamic networks. Here, one initializes the PageRanks of vertices with PageRanks obtained from the previous snapshot of the graph and runs the PageRank algorithm on all vertices. PageRanks obtained through this method will be at least as accurate as those obtained through the static algorithm. In this report, we refer to barrier-based Naive-dynamic (ND) PageRank as \textbf{\NaiWbar{}}.\ignore{Zhang et al. \cite{rank-zhang17} have explored the \textit{Naive-dynamic} approach in the hybrid CPU-GPU setting.} We also apply the ND approach to our improved lock-free PageRank, and refer to it as \textbf{\NaiBarf{}}. Similar to \StaBarf{}, it uses a flag vector $R_C$ to check for convergence. The psuedocode of \NaiWbar{} and \NaiBarf{} is given in Algorithms \ref{alg:with-barrier-naive-dynamic} and \ref{alg:barrier-free-naive-dynamic}\ignore{respectively}.

\subsubsection{Dynamic Traversal (DT) approach \cite{rank-desikan05, kim2015incremental, rank-giri20, rank-sahu22}}
\label{sec:about-traversal}

Proposed by Desikan et al. \cite{rank-desikan05}, this is a widely adopted approach for updating PageRank where one skips processing of vertices that have no impact on their PageRank due to the given batch update. For each edge deletion/insertion $(u, v)$ in the batch update, one marks all the vertices reachable from the vertex $u$ in the graph $G^{t-1}$ or the graph $G^t$ as affected, using Depth First Search (DFS) or Breadth First Search (BFS).\ignore{Giri et al. \cite{rank-giri20} have explored the \textit{Dynamic Traversal} approach in the hybrid CPU-GPU setting. On the other hand, Banerjee et al. \cite{rank-sahu22} have explored this approach in the CPU and GPU settings separately where they compute the ranks of vertices in topological order of strongly connected components (SCCs) to minimize unnecessary computation. They borrow this ordered processing of SCCs from the original static algorithm proposed by Garg et al. \cite{rank-garg16}.} We refer to barrier-based Dynamic Traversal (DT) PageRank as \textit{\TraWbar{}}. In a lock-free implementation of this approach, which we refer to as \textit{\TraBarf{}}, we allow any thread to start computing ranks as soon as it has finished marking affected vertices (based on edge updates assigned to it). This requires the use of a batch update checked flag $C$, for each source vertex, in addition to the flag vector $R_C$, used to check for convergence. Both \TraWbar{} and \TraBarf{} use a flag vector $V_A$, to indicate the vertices that have been marked as affected due to the current batch update. The psuedocode of \TraWbar{} and \TraBarf{} is given in Algorithms \ref{alg:with-barrier-dynamic-traversal} and \ref{alg:barrier-free-dynamic-traversal} respectively.

However, our experiments show that this approach cannot perform better than the ND approach for any batch size. The overhead of this approach due to several traversals needed to identify affected vertices limits the performance of this approach. Accordingly, we do not discuss this approach further.


%% file: 03-approach.tex
We have already discussed our lock-free implementations of Static and Naive-dynamic (ND) PageRank, termed \StaBarf{} and \NaiBarf{} respectively. Now, we turn our attention to our proposed Dynamic Frontier (DF) approach for updating PageRanks on dynamic graphs. This approach efficiently identifies and processes vertices whose PageRanks are likely to change due to a batch update, and it constitutes the main focus of our report.

\subsection{Our Dynamic Frontier (DF) approach}
\label{sec:frontier}

If a batch update $\Delta^{t-} \cup \Delta^{t+}$ is small compared to the total number of edges $|E|$, then it is expected that the PageRanks of only a few vertices change. Our proposed Dynamic Frontier (DF) approach incorporates this by identifying and processing affected vertices concurrently via an incremental process.

\input{src/fig-about-frontier}

\subsubsection{Explanation of the approach}
\label{sec:frontier-explanation}

Consider a batch update consisting of edge deletions $(u, v) \in \Delta^{t-}$ and insertions $(u, v) \in \Delta^{t+}$. We first initialize the PageRank of each vertex to that obtained in the previous snapshot of the graph.

\paragraph{Initial marking of affected vertex on edge deletion/insertion:}

For each edge deletion/insertion $(u, v)$, we initially mark the outgoing neighbors of the vertex $u$ in the previous graph $G^{t-1}$ and the current graph $G^t$ as affected\ignore{(lines \ref{alg:with-barrier--mark-begin}-\ref{alg:with-barrier--mark-end} in Algorithm \ref{alg:with-barrier}, and lines \ref{alg:barrier-free--mark-begin}-\ref{alg:barrier-free--mark-end} in Algorithm \ref{alg:barrier-free})}.

\paragraph{Incremental marking of affected vertices\ignore{upon change in rank of a given vertex}:}

Next, while performing PageRank computation\ignore{(lines \ref{alg:with-barrier--compute-begin}-\ref{alg:with-barrier--compute-end} in Algorithm \ref{alg:with-barrier}, and lines \ref{alg:barrier-free--compute-begin}-\ref{alg:barrier-free--compute-end} in Algorithm \ref{alg:barrier-free})}, if the PageRank of any affected vertex $v$ changes by more than the \textit{frontier tolerance} $\tau_f$ in an iteration, we mark its outgoing neighbors as affected\ignore{(lines \ref{alg:with-barrier--remark-begin}-\ref{alg:with-barrier--remark-end} in Algorithm \ref{alg:with-barrier}, and lines \ref{alg:barrier-free--remark-begin}-\ref{alg:barrier-free--remark-end} in Algorithm \ref{alg:barrier-free})}. This process of marking vertices as affected and processing such affected vertices continues in every iteration --- until convergence.

\subsubsection{A simple example}

Figure \ref{fig:about-frontier} shows an example of the DF approach. The original graph, shown in Figure \ref{fig:about-frontier-01} consists of $14$ vertices. Subsequently, Figure \ref{fig:about-frontier-02} shows a batch update applied to the graph involving the deletion of an edge from vertex $10$ to $11$ and insertion of an edge from vertex $7$ to $9$.

Following the batch update, we mark the outgoing neighbors of $7$ and $10$ as affected, i.e., $8$, $9$, and $11$ are marked as affected. Note that vertices $7$ and $10$ are not affected as they are a source of the change. We can now start the first iteration of PageRank.

In the first iteration, ranks of affected vertices are updated. Suppose that the change in rank of vertices $8$, $9$, and $11$ is observed to be greater than frontier tolerance $\tau_f$, shown with red border in Figure \ref{fig:about-frontier-03}. In response to this, the DF approach incrementally marks the out-neighbors of $8$, $9$, and $11$ as affected, i.e., $10$ and $12$.

During the second iteration, see Figure \ref{fig:about-frontier-04}, the PageRanks of affected vertices are again updated. Suppose that the change in PageRank of no vertex is observed to be greater than frontier tolerance $\tau_f$. In the subsequent iteration, the PageRanks of affected vertices are again updated. If the change in PageRank of each vertex is within iteration tolerance $\tau$, the PageRanks of vertices have converged, and the algorithm terminates.

\subsection{Our barrier-based DF PageRank (\FroWbar{})}
\label{sec:frontier-withbarrier}

We now describe\ignore{the implementation of} our barrier-based Dynamic Frontier (DF) PageRank, which we refer to as \FroWbar{}, given in Algorithm \ref{alg:with-barrier}. It takes as input the previous graph $G^{t-1}$ and the current graph $G^t$, edge deletions $\Delta^{t-}$ and insertions $\Delta^{t+}$ in the batch update, the previous PageRank vector $R^{t-1}$, and returns the updated PageRanks $R^t = R$.

In the algorithm, we begin by initializing the PageRank vectors $R$ and $R_{new}$ for consecutive iterations, with the previous PageRank vector $R^{t-1}$ (line \ref{alg:with-barrier-initialize}). As mentioned in Section \ref{sec:withbarrier}, we perform a synchronous PageRank computation with \FroWbar{} using two PageRank vectors $R$, and $R_{new}$. Next, for each edge in the batch update, i.e., $\Delta^{t-}$ or $\Delta^{t+}$, we initially mark all out-neighbors of the source vertex $u$ in both $G^{t-1}$ or $G^t$ as affected in parallel (lines \ref{alg:with-barrier--mark-begin}-\ref{alg:with-barrier--mark-end}). To track vertices that are marked as affected due to the current batch update $\Delta^{t-} \cup \Delta^{t+}$, we use an 8-bit integer based flag vector.

We then iteratively compute the PageRank $r$ for each affected vertex $v$ (lines \ref{alg:with-barrier--compute-begin}-\ref{alg:with-barrier--compute-end}). This computation is performed in parallel, considering the incoming edges $G^t.in(v)$. We then check if the change in PageRank $\Delta r$ exceeds the frontier tolerance $\tau_f$, and if so, we mark out-neighbors of $v$ as affected. There is an implicit iteration barrier at the end of the loop, where threads synchronize after computing PageRanks in parallel. Next, in parallel, we compute the maximum error $\Delta R$ between the PageRanks obtained in the previous $R$ and the current iteration $R_{new}$ with $L_\infty$-norm. We then swap $R_{new}$ and $R$ to set up the next iteration. The iteration (lines \ref{alg:with-barrier--iterations-begin}-\ref{alg:with-barrier--iterations-end}) continues until either the maximum change in PageRanks $\Delta R$ falls below the iteration tolerance $\tau$, or the maximum number of iterations $MAX\_ITERATIONS$ is reached. At the end, the algorithm returns the final PageRank vector $R$ (line \ref{alg:with-barrier--return}).

\input{src/alg-with-barrier}
\input{src/alg-barrier-free}

\subsection{Our lock-free DF PageRank (\FroBarf{})}
\label{sec:frontier-barrierfree}

We now explain the working of our lock-free algorithm, \FroBarf{}. Note that converting \FroWbar{} (Algorithm \ref{alg:with-barrier}) into a lock-free algorithm is not just a matter of adding OpenMP \texttt{nowait} directive to remove the implicit iteration barrier (lines \ref{alg:with-barrier--implicit-barrier-mark}, \ref{alg:with-barrier--implicit-barrier-compute} in Algorithm \ref{alg:with-barrier}). Doing so causes either the program to abruptly crash, or return incorrect PageRanks. Hence, it must be used with caution. In addition, in order to make \FroWbar{} lock-free, one cannot make use of the $swap()$ operation (line \ref{alg:with-barrier--swap}), as this needs to happen on a single thread at the end of each iteration --- but with a lock-free algorithm, the threads must be operating independently. Instead, one must thus switch to an asynchronous version of PageRank algorithm, which uses a single PageRank vector $R$ (line \ref{alg:barrier-free--initialize} in Algorithm \ref{alg:barrier-free}). Again, as the thread needs to be operating independently, one must move the convergence check to each thread, and each thread must share the convergence status of each vertex/chunk with the other threads. This can be achieved through the use of a per-vertex/per-chunk \textit{converged} flag $R_C$. The computation stops when the PageRanks of all vertices have converged. Note that, unlike with \FroWbar{}, it is not just desirable, but necessary, to use OpenMP's \textit{dynamic} schedule in PageRank computation loop\ignore{of each iteration} (line \ref{alg:barrier-free--compute-begin}) to achieve fault tolerance.

\ignore{However, in contrast to \FroWbar{}, with \FroBarf{} we perform an asynchronous PageRank computation using a single PageRank vector $R$. Further, we remove implicit barriers by encapsulating both steps (i.e., marking an initial set of affected vertices, and performing PageRank computation) in a common parallel block (lines \ref{alg:barrier-free--parallel-begin}-\ref{alg:barrier-free--parallel-end}).}

Further, note that \FroWbar{} has two phases, the initial marking of affected vertices (lines \ref{alg:with-barrier--mark-begin}-\ref{alg:with-barrier--mark-end} in Algorithm \ref{alg:with-barrier}), and the incremental marking, processing, and convergence detection of affected vertices (lines \ref{alg:with-barrier--compute-begin}-\ref{alg:with-barrier--compute-end} in Algorithm \ref{alg:with-barrier}). Composing these two phases into a fault-tolerant lock-free algorithm requires threads to help each other. This ensures that if some thread is slow in completing its work, then other threads will come forward to complete its work. We achieve this with the use of a top-level parallel block (lines \ref{alg:barrier-free--parallel-begin}-\ref{alg:barrier-free--parallel-end} in Algorithm \ref{alg:barrier-free}), use of \texttt{nowait} clause in the initial marking phase of the algorithm (line \ref{alg:barrier-free--mark-check-start}), and with the use of a \textit{processed} flag vector $C$ to indicate that the algorithms appropriately processes each edge in the batch update before the PageRank computation begins\ignore{(line \ref{alg:barrier-free-mark-done})}.

\ignore{\subsubsection{Our solution}}

We now explain the pseudocode of \FroBarf{}, which is given in Algorithm \ref{alg:barrier-free}. Similar to \FroWbar{}, it takes as input the previous graph $G^{t-1}$ and the current graph $G^t$, edge deletions $\Delta^{t-}$ and insertions $\Delta^{t+}$ in the batch update, the previous PageRank vector $R^{t-1}$, and returns the updated PageRank vector $R$.

In the algorithm, we begin by initializing flag vectors $C$ and $R_C$ to indicate that none of the source vertices $u$ from the batch update $(u, v) \in \Delta^{t-} \cup \Delta^{t+}$ have been checked, and that PageRanks of all vertices converged, respectively (line \ref{alg:barrier-free--initialize}). The vector $C$ is used to identify all the source vertices in the batch update $\Delta^t$. Outgoing neighbors of such vertices must have their ranks recomputed, and thus must be marked as affected, as they are immediately impacted by edges in the batch update. The vector $R_C$ indicates if the PageRanks for each vertex has converged or not. Next, we initialize the current PageRank vector $R$ with the previous PageRank vector $R^{t-1}$ (also line \ref{alg:barrier-free--initialize}). We then start a top-level parallel block (lines \ref{alg:barrier-free--parallel-begin}-\ref{alg:barrier-free--parallel-end}) to encapsulate the two main phases of \FroBarf{}, i.e., the initial marking of affected vertices (lines \ref{alg:barrier-free--mark-begin}-\ref{alg:barrier-free--mark-end}), and the incremental marking of affected vertices and computation of PageRanks (lines \ref{alg:barrier-free--compute-begin}-\ref{alg:barrier-free--compute-end}).

\paragraph{Initial marking of affected vertices}

To mark the initial set of affected vertices (lines \ref{alg:barrier-free--mark-begin}-\ref{alg:barrier-free--mark-end}), we iterate over the edge deletions and insertions $(u, v) \in \Delta^{t-} \cup \Delta^{t+}$ in parallel using OpenMP's \texttt{nowait} directive. If a source vertex $u$ has not been processed ($C[u] = 0$), we mark all its out-neighbors in both $G^{t-1}$ and $G^t$ as affected ($V_A[v'] = 1$ for each outgoing neighbor $v'$ of $u$), mark them as needing PageRank update ($R_C[v'] = 1$), and also update $C[u]$ to indicate that the source vertex $u$ has been checked. We use an 8-bit integer vector to keep track of the vertices identified as affected by the current batch update $\Delta^{t-} \cup \Delta^{t+}$, as with \FroWbar{}. Flag vectors $C$ and $R_C$ are also 8-bit integer vectors.

Due to the barrier-free nature of the algorithm, it is possible for a thread to start the next phase of the algorithm (i.e. computing PageRanks in lines \ref{alg:barrier-free--compute-begin}-\ref{alg:barrier-free--compute-end}), before all the affected vertices have been marked. The flag vector $C$ enables us to avoid this scenario, and to allow threads to help one another. A value of $1$ in the vector $C$ for $u$ ($C[u]=1$) indicates that a vertex $u$ in the batch update $(u, v) \in \Delta^{t-} \cup \Delta^{t+}$ has been checked by some thread in the initial marking phase of the DF approach. Now, once all the out-neighbors of $u$ have been marked as affected by setting corresponding entries in $V_A$, the entry $C[u]$ is set to $1$ in line \ref{alg:barrier-free-mark-done}. After a thread has finished the initial marking phase based on dynamically assigned edge updates, it verifies whether all batch updates have been reviewed by the other threads, i.e., $C[u]=1\ \forall\ (u, v) \in \Delta^{t-} \cup \Delta^{t+}$ (lines \ref{alg:barrier-free--mark-check-begin}-\ref{alg:barrier-free--mark-check-end} in Algorithm \ref{alg:barrier-free}). If not, the thread continues to check edge updates that have not been reviewed and marks appropriate vertices as affected. This is crucial in handling delays or crashes experienced by threads and helps achieve lock-freedom in the initial marking of affected vertices. As threads are helping each other, a thread can skip duplicate work by checking if $C[u]$ is already set to $1$ for a given edge update in line \ref{alg:barrier-free-mark-skip}.

\paragraph{Incremental marking and processing of affected vertices}

When all edge updates have been reviewed, threads can independently proceed to compute PageRanks (second phase of \FroBarf{}). In this phase (lines \ref{alg:barrier-free--compute-begin}-\ref{alg:barrier-free--compute-end}), each thread iterates over the affected vertices in parallel by dynamically picking up a chunk of vertices to process, while using OpenMP's \texttt{nowait} directive. Each thread first computes the new PageRank $r$ of each affected vertex $v$ based on the PageRank algorithm, and updates it in-place on the common PageRank vector $R$. Next, it calculates the change in PageRank $\Delta r$ for vertex $v$. It then check if $\Delta r$ exceeds the frontier tolerance $\tau_f$, and if so, it marks each out-neighbor $v'$ of $v$ as affected and needing PageRank update by setting $R_C[v']$ to $1$. Further, if the change in PageRank $\Delta r$ for vertex $v$ is within the iteration tolerance $\tau$, the thread marks it as converged by setting $R_C[v]$ to $0$ (line \ref{alg:barrier-free--converged-vertex}). The flag vector $R_C$ (also an 8-bit integer vector) is employed as a per-vertex convergence flag, with a $1$ ($R_C[v]=1$) indicating that the PageRank of a vertex $v$ has not yet converged. In other words, $R_C[v]=1$ implies that $\Delta r$ (which represents the change in PageRank of a single vertex) for vertex $v$ is greater than the specified (iteration) tolerance $\tau$. Our use of this flag vector allows independently running threads to share information on the convergence status of the rank of each vertex in the graph. Alternatively, one may use a per-chunk converged flag for even faster detection of convergence. Each thread then check if all vertices have converged. If all vertices have converged, i.e., $R_C[v]=0\ \forall\ v \in V^t$ (line \ref{alg:barrier-free--converged-all}), the thread breaks from the loop, and returning the final PageRank vector $R$ from the main thread.

\subsection{Lock-freedom and fault tolerance of \FroBarf{}}
\label{sec:frontier-tolerate}

We now explain how \FroBarf{} is lock-free. First consider the initial marking phase of \FroBarf{} in lines \ref{alg:barrier-free--mark-begin}-\ref{alg:barrier-free--mark-end} of Algorithm \ref{alg:barrier-free}. Here, a thread $th_i$ picks up an edge $(u, v)$ from the batch update for processing. If a thread $th_s$ stalls --- for a certain period of time, or indefinitely --- after selecting a chunk of edges $\Delta_s \subseteq \Delta^{t-} \cup \Delta^{t+}$ for processing with \FroBarf{}, other threads will not advance to the next phase as they would observe unmarked out-neighbors in $\Delta_s$ via the $C$ flag vector. This triggers a race among threads to process $\Delta_s$, which may lead to parallel processing by multiple threads. However, the idempotent nature of the operations ensures that this race does not affect the final result. Once $\Delta_s$ is completely processed, all remaining threads progress to the next phase which is achieved without the use of barriers. As at least one thread progresses in the presence of faults, this phase is both lock-free and fault tolerant.

\ignore{We now analyze \FroBarf{} to explain how it achieves fault tolerance, and is lock-free. First consider the loop in lines \ref{alg:barrier-free--mark-check-start}-\ref{alg:barrier-free--mark-check-end} of Algorithm \ref{alg:barrier-free}. Here, a thread $th_i$ picks up an edge $(u, v)$ from the batch update for processing. It is possible for another thread $th_j$ to also pick up an edge with the same source vertex $u$ simultaneously. Thread $th_i$ upon verifying that $C[u]$ is $0$, proceeds to update the flag vectors $V_A$ and $R_C$ for each outgoing neighbour of $u$. Thread $th_j$ working on the same source vertex $u$ concurrently, completes the same work without interfering with $th_i$. Thus, the threads $th_i$ and $th_j$ work in an idempotent manner. Further, if $th_i$ experiences a slowdown or crashes inside this loop, then another thread $th_k$ can mark the outgoing neighbors of $u$ as affected, either in the current or the next iteration. This ensures that all edges in the batch update are processed in a \lf manner.}

Now consider the subsequent phase of the algorithm (lines \ref{alg:barrier-free--compute-begin}-\ref{alg:barrier-free--compute-end}). If a thread $th_s$ becomes stuck during the incremental marking and PageRank computation phase, having selected a chunk of vertices $V_s \subseteq V^t$ for processing, no other thread will terminate the algorithm. They will observe via the $R_C$ flag vector that the PageRanks of vertices in $V_s$ have not converged yet. Consequently, they will continue processing all vertices, including those in $V_s$, in the next iteration until all vertex PageRanks have converged. It is possible that $th_s$ resumes execution later and updates a vertex's PageRank $v_s$ in $V_s$ after it has already been modified by other threads. However, given the nature of PageRank computation, work done by $th_s$ does not affect the correctness and PageRanks will eventually converge as other threads can fix/update the PageRank of $v_s$ in the next iteration.\ignore{Despite non-idempotent operations, PageRanks will eventually converge due to the nature of PageRank algorithm.} Once all PageRanks have converged, the remaining threads will terminate, and the final result will be returned. As vertices in $V_s$ are updated in the next iteration by some active thread, and progress is ensured by at least one thread.\ignore{Finally, both phases of \FroBarf{} and encapsulated in a common parallel block (lines \ref{alg:barrier-free--parallel-begin}-\ref{alg:barrier-free--parallel-end}).} Thus, \FroBarf{} is\ignore{barrier-free is both} lock-free and fault-tolerant.

\ignore{Our algorithms converge as we apply the commonly-used power-iteration algorithm, where PageRanks of vertices are updated in each iteration. With Barrier-free PageRank, in case of a thread crash or sleep, the PageRanks will be updated in the next iteration. The algorithm terminates when PageRanks converge or $MAX\_ITERATIONS$ is reached --- same as with fault-free PageRank.}

\ignore{By employing a common top-level parallel block for the initial marking of affected vertices and the updating of ranks until convergence, utilizing OpenMP's dynamic loop scheduling along with \texttt{nowait} directive, using a single shared rank vector, making use of shared flag vectors, and utilizing flag vectors $C$ and $R_C$ to verify the completion of the two phases of the algorithm (i.e., initial marking of affected vertices, and the updating of ranks), \FroBarf{} achieves overall \lfdm and \textit{fault tolerance}.}

\subsection{Determining Frontier tolerance ($\tau_f$)}

Our experiments show that a frontier tolerance of $\tau_f = \tau/1000$, where $\tau$ is the iteration tolerance, provides a good speedup with a maximum error of $10^{-9}$ at a batch size of $10^{-4}|E|$, compared to a mean error of $5\times10^{-10}$ for \NaiWbar{} and \NaiBarf{} --- with respect to PageRanks obtained from reference PageRank (see Section \ref{sec:measurement}). Thus, if an edge in the batch update affects the PageRank of vertex $v$ (beyond a threshold amount), directly or indirectly, all its outgoing neighbors $v' \in G^t.out(u)$ will also be marked as affected, as they are likely to have a change in PageRank as well.

%% file: src/fig-about-frontier.tex
\begin{figure*}[hbtp]
  \centering
  \subfigure[Initial graph]{
    \label{fig:about-frontier-01}
    \includegraphics[width=0.23\linewidth]{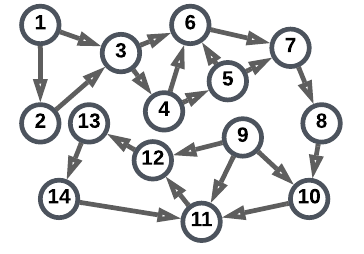}
  }
  \subfigure[Marking affected (initial)]{
    \label{fig:about-frontier-02}
    \includegraphics[width=0.23\linewidth]{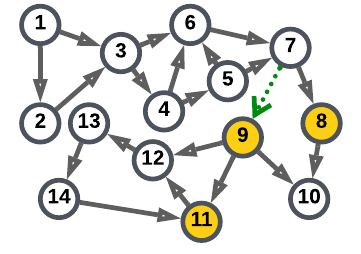}
  }
  \subfigure[After first iteration]{
    \label{fig:about-frontier-03}
    \includegraphics[width=0.23\linewidth]{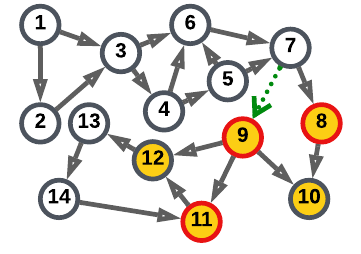}
  }
  \subfigure[After second iteration]{
    \label{fig:about-frontier-04}
    \includegraphics[width=0.23\linewidth]{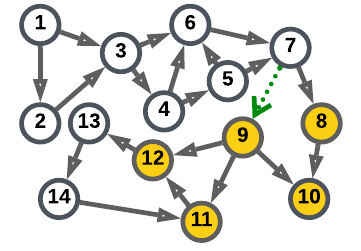}
  } \\[-1ex]
  \caption{An example explaining the \textit{Dynamic Frontier (DF)} approach. The original graph with 14 vertices undergoes a batch update: deleting the edge from vertex $10$ to $11$ and inserting an edge from vertex $7$ to $9$. Affected vertices are shown with yellow fill, and vertices with a change in rank greater than frontier tolerance $\tau_f$ are shown with red border.}
  \label{fig:about-frontier}
\end{figure*}

%% file: src/alg-with-barrier.tex
\begin{algorithm}[!hbt]
\caption{Our Barrier-based Dynamic Frontier PageRank\ignore{($\text{DF}_{\text{BB}}$)}.}
\label{alg:with-barrier}
\begin{algorithmic}[1]
\Require{$G^{t-1}, G^{t}$: Previous, current input graph}
\Require{$\Delta^{t-}, \Delta^{t+}$: Edge deletions and insertions}
\Require{$R^{t-1}$: Previous rank vector}
\Ensure{$V_A[u]$: Is vertex $u$ affected due to current batch update?}
\Ensure{$R, R_{new}$: Rank vectors in current, previous iteration}
\Ensure{$\Delta R$: $L_\infty$-norm between ranks in consecutive iterations}
\Ensure{$\Delta r$: Change in rank of a vertex}
\Ensure{$\tau, \tau_f$: Iteration, frontier tolerance}
\Ensure{$\alpha$: Damping factor}

\Statex

\Function{$\text{DF}_{\text{BB}}$}{$G^{t-1}, G^t, \Delta^{t-}, \Delta^{t+}, R^{t-1}$}
  \State $R_{new} \gets R \gets R^{t-1}$
  \State $//$ Mark initial affected \label{alg:with-barrier-initialize}
  \ForAll{$(u, v) \in \Delta^{t-} \cup \Delta^{t+} \textbf{in parallel}$} \label{alg:with-barrier--mark-begin}
    \ForAll{$v' \in (G^{t-1} \cup G^t).out(u)$}
    \State Mark $v'$ as affected
    \EndFor
  \EndFor \label{alg:with-barrier--mark-end}
  \State \textbf{wait for all threads} (implicit barrier) \label{alg:with-barrier--implicit-barrier-mark}
  \ForAll{$i \in [0 .. MAX\_ITERATIONS)$} \label{alg:with-barrier--iterations-begin}
    \ForAll{affected $v \in V^t$ \textbf{in parallel}} \label{alg:with-barrier--compute-begin}
      \State $r \gets (1 - \alpha)/|V^t|$
      \ForAll{$u \in G^t.in(v)$}
        \State $r \gets r + \alpha \times R[u] / |G^t.out(u)|$
      \EndFor
      \State $\Delta r \gets |r - R[v]|$ \textbf{;} $R_{new}[v] \gets r$
      \State $//$ Is rank change $>$ frontier tolerance?
      \If{$\Delta r > \tau_f$} \label{alg:with-barrier--remark-begin}
        \ForAll{$v' \in G^t.out(v)$}
          \State Mark $v'$ as affected
        \EndFor
      \EndIf \label{alg:with-barrier--remark-end}
    \EndFor \label{alg:with-barrier--compute-end}
    \State \textbf{wait for all threads} (implicit barrier) \label{alg:with-barrier--implicit-barrier-compute}
    \State $\Delta R \gets l\infty Norm(R, R_{new})$ \textbf{in parallel}
    \State $swap(R_{new}, R)$ \label{alg:with-barrier--swap}
    \State $//$ Ranks converged?
    \If{$\Delta R \le \tau$} \textbf{break}
    \EndIf
  \EndFor \label{alg:with-barrier--iterations-end}
  \State \ReturnInline{$R$} \label{alg:with-barrier--return}
\EndFunction
\end{algorithmic}
\end{algorithm}

%% file: src/alg-barrier-free.tex
\begin{algorithm}[!hbt]
\caption{Our Lock-free Dynamic Frontier PageRank\ignore{($\text{DF}_{\text{LF}}$)}.}
\label{alg:barrier-free}
\begin{algorithmic}[1]
\Require{$G^{t-1}, G^{t}$: Previous, current input graph}
\Require{$\Delta^{t-}, \Delta^{t+}$: Edge deletions and insertions}
\Require{$R^{t-1}$: Previous rank vector}
\Ensure{$V_A[u]$: Is vertex $u$ affected due to batch update?}
\Ensure{$C[u]$: Has vertex $u$ from batch update been checked?}
\Ensure{$R_C[u]$: Has the rank of vertex $u$ not yet converged?}
\Ensure{$R, R_{new}$: Rank vectors in current, previous iteration}
\Ensure{$\Delta R$: $L_\infty$-norm between ranks in consecutive iterations}
\Ensure{$\Delta r$: Change in rank of a vertex}
\Ensure{$\tau, \tau_f$: Iteration, frontier tolerance}
\Ensure{$\alpha$: Damping factor}

\Statex

\Function{$\text{DF}_{\text{LF}}$}{$G^{t-1}, G^t, \Delta^{t-}, \Delta^{t+}, R^{t-1}$}
  \State $C \gets R_C \gets \{0\}$ \textbf{;} $R \gets R^{t-1}$ \label{alg:barrier-free--initialize}
  \ForAll{threads \textbf{in parallel}} \label{alg:barrier-free--parallel-begin}
    \State $//$ Mark initial affected
    \While{$true$} \label{alg:barrier-free--mark-begin}
    \State $//$ For loop with no-wait
     \ForAll{$(u, v) \in \Delta^{t-} \cup \Delta^{t+}$ \textbf{dyn. sched.}} \label{alg:barrier-free--mark-check-start}
        \State $//$ Not marked yet?
         \If{$C[u] = 0$} \label{alg:barrier-free-mark-skip} 
            \ForAll{$v' \in (G^{t-1} \cup G^t).out(u)$}
            \State Mark $v'$ as affected
            \State $R_C[v'] \gets 1$
          \EndFor
        \EndIf
        \State $C[u] \gets 1$ \label{alg:barrier-free-mark-done}
      \EndFor
      \State $//$ All marked?
      \If{$C[u]=1\ \forall\ (u, v) \in \Delta^{t-} \cup \Delta^{t+}$} \label{alg:barrier-free--mark-check-begin}
        \State \textbf{break}
      \EndIf \label{alg:barrier-free--mark-check-end}
    \EndWhile \label{alg:barrier-free--mark-end}
    \ForAll{$i \in [0 .. MAX\_ITERATIONS)$} \label{alg:barrier-free--iterations-begin}
      \State $//$ For loop with no-wait
      \ForAll{affected $v \in V^t$ \textbf{dyn. sched.}} \label{alg:barrier-free--compute-begin}
        \State $r \gets (1 - \alpha)/|V^t|$
        \ForAll{$u \in G^t.in(v)$}
          \State $r \gets r + \alpha \times R[u] / |G^t.out(u)|$ \label{alg:barrier-free--calculate-rank}
        \EndFor
        \State $\Delta r \gets |r - R[v]|$ \textbf{;} $R[v] \gets r$
        \State $//$ Is rank change $>$ frontier tolerance?
        \If{$\Delta r > \tau_f$} \label{alg:barrier-free--remark-begin}
          \ForAll{$v' \in G^t.out(v)$}
            \State Mark $v'$ as affected
            \State $R_C[v'] \gets 1$
          \EndFor
        \EndIf \label{alg:barrier-free--remark-end}
        \If{$\Delta r \le \tau$} $R_C[v] \gets 0$ \label{alg:barrier-free--converged-vertex}
        \EndIf
      \EndFor \label{alg:barrier-free--compute-end}
      \State $//$ Ranks converged?
      \If{$R_C[v] = 0\ \forall\ v \in V^t$} \textbf{break} \label{alg:barrier-free--converged-all}
      \EndIf
    \EndFor \label{alg:barrier-free--iterations-end}
  \EndFor \label{alg:barrier-free--parallel-end}
  \State \ReturnInline{$R$}
\EndFunction
\end{algorithmic}
\end{algorithm}

%% file: 04-evaluation.tex
\subsection{Experimental setup}
\label{sec:setup}

\subsubsection{System}
\label{sec:system}

We experiment on a system consisting of a $64$-core x86-based AMD EPYC-7742 processor running at $2.25$ GHz. Each core has an L1 cache of $4$ MB, an L2 cache of $32$ MB, and a shared L3 cache of $256$ MB. The server has $512$ GB DDR4 system memory, and runs on Ubuntu $20.04$. We use GCC $9.4$ and OpenMP $5.0$.

\subsubsection{Configuration}
\label{sec:configuration}

We use 32-bit integers for vertex ids and 64-bit floats for computing and storing PageRanks. To mark vertices as affected, we use an 8-bit integer vector. The rank computation uses OpenMP's \textit{dynamic schedule} with a chunk size of $2048$, which provides dynamic work-balancing among threads. We use a damping factor of $\alpha = 0.85$ \cite{rank-langville06}, an iteration tolerance of $\tau = 10^{-10}$ \cite{rank-dubey22} using $L\infty$-norm \cite{rank-plimpton11}, and limit the \texttt{MAX\_ITERATIONS} performed to $500$ \cite{nvgraph} for all experiments. Unless specified otherwise, we run all experiments with $64$ threads (one thread per core).

\subsubsection{Dataset}
\label{sec:dataset}

We test on two large real-world dynamic graphs from the Stanford Large Network Dataset Collection \cite{snap14}, shown in Table \ref{tab:dataset-temporal}. Here, the number of vertices ranges from $1.14$ million to $2.60$ million, and the number of temporal edges varies from $7.83$ million to $63.4$ million. The presence of dead ends (vertices with no out-links) introduces a global teleport rank contribution that must be computed every iteration. We eliminate this overhead by adding self-loops to all the vertices in the graph \cite{rank-andersen07, rank-langville06}.

In addition, we also use four classes of graphs from the SuiteSparse Matrix Collection \cite{suite19} in our experiments, shown in Table \ref{tab:dataset}. The number of vertices in these graphs varies from $3.07$ million to $214$ million, and the number of edges varies from $37.4$ million to $1.98$ billion. For undirected graphs in Table \ref{tab:dataset}, we add two directed edges for each edge in the undirected graph. As above, we remove dead ends by adding self-loops to all vertices in each graph.

\input{src/tab-dataset-temporal}
\input{src/tab-dataset}

\subsubsection{Batch Generation}
\label{sec:batch-generation}

The real-world dynamic graphs in Table \ref{tab:dataset-temporal} are datasets containing edge insertions paired with\ignore{UNIX} timestamps. To conduct experiments on these graphs in the batch dynamic setting, we initially load $90\%$ of the original graph. Then, we apply batch updates by loading and applying\ignore{$B$} edges to the original graph as a batch update, one after the other, rather than limiting updates based on timestamps. The batch update size is either $10^{-4}|E_T|$ or $10^{-3}|E_T|$, resulting in $1000$ batch updates with a size of $10^{-4}|E_T|$ and $100$ batch updates with a size of $10^{-3}|E_T|$. This process allows us to read the entire real-world dynamic graph. Since the graphs in Table \ref{tab:dataset-temporal} only involve edge insertions, our experiments on these graphs solely focus on edge insertions.

For graphs in Table \ref{tab:dataset}, we take each graph and generate a random batch update consisting of an equal mix of edge deletions and insertions. To prepare the set of edges deleted, we delete each existing edge with a uniform probability. We prepare the set of edges to insert by choosing non-connected pairs of vertices with equal probability. For the sake of simplicity, we ensure that no new vertices are added to or removed from the graph. We measure the batch size as a fraction of the total number of edges in the original graph and adjust it from $10^{-8}$ to $0.1$ (i.e., $10^{-8}|E|$ to $0.1|E|$). For averaging, we generate multiple batches for each batch size. Along with each batch update, we add self-loops to all vertices.

\subsubsection{Measurement}
\label{sec:measurement}

We measure the time taken by each approach on the graph entirely, including any preprocessing cost and the time taken to detect convergence --- but exclude time taken for memory allocation and de-allocation. The average time taken by a given method at a given batch size is obtained by taking the geometric mean of time taken for the same batch size for each of the different input graphs. Accordingly, the average speedup is the ratio of these. Further, we measure the error/accuracy of a given approach by measuring the $L_\infty$-norm of the PageRanks produced with respect to PageRanks obtained from a reference barrier-based Static PageRank run on the updated graph with a very low tolerance of $\tau = 10^{-100}$, limited to $500$ iterations.

\subsubsection{Fault simulation}
\label{sec:fault-simulation}

We simulate a random thread delay such that it can occur after computing the rank of any vertex in an iteration with a certain probability. This random thread delay affects \textit{all} threads uniformly. We similarly simulate a random thread crash by setting a per-thread \texttt{crashed} flag, which signals that particular thread to stop its execution deterministically (crash-stop model).

\subsection{Performance in the Absence of Faults}
\label{sec:fault-none}

\subsubsection{Performance on Real-world dynamic graphs}

We begin by analyzing the performance of \FroWbar{} and \FroBarf{} on real-world dynamic graphs, as detailed in Table \ref{tab:dataset-temporal}, utilizing batch updates of size $10^{-4}|E_T|$ and $10^{-3}|E_T|$. We compare the results of \FroWbar{} and \FroBarf{}  with respect to \StaWbar{}, \StaBarf{}, \NaiWbar{}, and \NaiBarf{}.

In Figure \ref{fig:temporal}, the results illustrate that our proposed \FroBarf{} achieves an average speedup of $3.8\times$, $3.2\times$, $4.5\times$, and $2.5\times$ over \StaWbar{}, \NaiWbar{}, \StaBarf{}, and \NaiBarf{}, respectively. Thus, the DF approach performs better on real-world dynamic graphs. Furthermore, \FroBarf{} demonstrates an average speedup of $1.6\times$ over \FroWbar{}, indicating that a lock-free computation of dynamic PageRank proves more efficient than a barrier-based approach. The increased efficiency is due to reduced wait times\ignore{, no copy-overhead for non-updated ranks,} and faster convergence of asynchronous dynamic PageRank.

\input{src/fig-temporal}

\subsubsection{Performance on Large graphs}

We now study the performance of \FroWbar{} and \FroBarf{} on large\ignore{(static)} graphs with\ignore{generated} random batch updates of size $10^{-8} |E|$ to $0.1 |E|$, in the absence of random thread delays or crashes, and compare them with \StaWbar{}, \StaBarf{}, \NaiWbar{}, and \NaiBarf{}. Figure \ref{fig:no-failure} plots the runtime of the six mentioned approaches. In the figure, solid lines represent the runtime of the barrier-free algorithms and dashed lines represent the runtime of the corresponding algorithms using a barrier. In addition to the runtime of the above algorithms on each instance from Table \ref{tab:dataset} shown in Figure \ref{fig:no-failure-all}, the mean runtime of each algorithm is shown in Figure \ref{fig:no-failure-am-time}. The labels on the line corresponding to \StaBarf{} and \NaiBarf{} indicate the speedup of \FroBarf{} over the respective algorithm.

From Figures \ref{fig:no-failure-all} and \ref{fig:no-failure-am-time}, we observe that our proposed \FroBarf{} is on average $12.6\times$, $5.4\times$, $12.0\times$, and $4.6\times$ is faster than \StaWbar{}, \NaiWbar{}, \StaBarf{}, and \NaiBarf{} until a batch size of $10^{-3} |E|$. From a batch size of $10^{-3} |E|$ (a batch update consisting of more than a million edge deletions/insertions on a billion-edge graph), the performance drops below \NaiBarf{} and \StaBarf{}. Such large batch sizes result in nearly all\ignore{of the} vertices getting marked as affected, and hence the performance drops. Note in Figure \ref{fig:no-failure-all} that \FroBarf{} performs well on road networks and protein k-mer graphs (sparse), but poorly on social networks (dense).\ignore{This seems to be associated to sparsity of the graphs.} Moreover, \FroBarf{} achieves a $1.6\times$ speedup over \FroWbar{} (as with real-world dynamic graphs).

The error with respect to reference PageRank for \FroBarf{} (as well as \FroWbar{}), as shown in Figure \ref{fig:no-failure-am-error}, starts increasing from $5\times10^{-10}$ to $9\times10^{-10}$ at a batch size of $10^{-6} |E|$ to $10^{-4} |E|$, and then drops back at a batch size of $10^{-2} |E|$. The Dynamic Frontier (DF) approach selects outgoing neighbors of vertices whose PageRank changes exceed a frontier tolerance of $\tau_f = \tau/1000$ (where iteration tolerance $\tau = 10^{-10}$). The error decreases due to more vertices being marked as affected with larger batch sizes. With \FroBarf{}, the error remains within the acceptable range of $[0, 10^{-9})$ for $\tau = 10^{-10}$. Our lock-free approach introduces no extra overhead in fault-free scenarios.

\subsubsection{Stability}
\label{sec:stability}

To measure the stability of our algorithms, we generate random batch updates of size $10^{-8} |E|$ to $0.1 |E|$ consisting purely of edge deletions, update PageRanks of vertices, insert the same edges back, and again update the PageRanks of vertices. Finally, we calculate $L_\infty$-norm of the difference in PageRanks obtained after these two batch updates with the PageRanks in the original graph. Ideally, the $L_\infty$-norm should be $0$.

\ignore{The mean $L\infty$-norm of ranks obtained from \StaBarf{}, \NaiBarf{} and \FroBarf{} after the deletions-only and insertions-only batch updates with respect to ranks obtained \StaBarf{} original graph is show in Figure \ref{fig:stability}. A similar comparison of \StaWbar{}, \NaiWbar{}, and \FroWbar{} is also shown.}

We observe that \FroWbar{} and \FroBarf{} have a maximum error of $5.7\times10^{-10}$ and $4.6\times10^{-10}$ respectively with the original PageRanks across all batch sizes, while \NaiWbar{} and \NaiBarf{} have a maximum error of $5.7\times10^{-10}$ and $4.6\times10^{-10}$ respectively, according to the $L_\infty$-norm. Thus, our DF approach is stable.

\ignore{\input{src/fig-stability}}

\subsubsection{Strong scaling}
\label{sec:scaling}

We next study the strong-scaling behavior of \FroWbar{} and \FroBarf{} on batch updates of a fixed size of $10^{-4} |E|$ in the absence of faults. Here, we measure the speedup of each algorithm with an increasing number of threads from $1$ to $64$ in multiples of $2$ with respect to a single-threaded execution of the algorithm.\ignore{We additionally compare \StaWbar{}, \StaBarf{}, \NaiWbar{}, and \NaiBarf{}.}

We observe from Figure \ref{fig:strong-scaling} that both algorithms exhibit a good speedup with increasing number of threads. With $32$ threads, \FroBarf{} offers a speedup of $19.5\times$ over a single thread, while \FroWbar{} offers a speedup of $14.4\times$ over a single thread. At $64$ threads, \FroBarf{} and \FroWbar{} are impacted by NUMA effects, and offer speedups of only $21.3\times$ and $14.5\times$ respectively.\ignore{Thus, \FroBarf{} offers good scaling performance.}

\input{src/fig-strong-scaling}
\input{src/fig-fault-none}

\subsection{Performance under Random Thread Delays}
\label{sec:fault-sleep}

We now study the behaviour of \FroBarf{} in the presence of random thread delays\ignore{which is common in asynchronous systems}. We simulate these via thread sleep, with probabilities of $10^{-9}|V|$ to $10^{-6}|V|$ per iteration. This simulated delay is equally likely for each thread. On a graph with $10$ million vertices, this amounts to an average of $0.01$ to $10$ thread sleeps per iteration (when all chunks have been picked up the running threads).\ignore{Our experiments show that if the duration of each random thread delay is small relative to the average iteration time ($100$ ms), the impact on algorithm's running time is minimal.} We try three different delays, $50$ ms, $100$ ms, and $200$ ms, which are sizeable relative to the iteration time.\ignore{In addition, JVM's garbage pause is set to a soft limit of $200$ ms \cite{jvm-beckwith13}.} We prepare a batch update of size $10^{-4} |E|$ for each graph, and compare \FroWbar{} and \FroBarf{}. We plot the runtime of both approaches in Figure \ref{fig:fault-sleep}. Here, solid lines represent the run time of \FroBarf{} on a batch size of $10^{-4} |E|$, and in the presence of random thread delays with probabilities of $10^{-9}|V|$ to $10^{-6}|V|$ per iteration. Dashed lines represent the run time of the corresponding with \FroWbar{}.

\input{src/fig-fault-sleep}
\input{src/fig-fault-crash}

We observe in Figures \ref{fig:uniform-sleep-all} and \ref{fig:uniform-sleep-am-time} that \FroWbar{} is affected by an increasing delay probability. In contrast, \FroBarf{} is minimally affected and outperforms \FroWbar{} by a significant margin when thread sleeps are more common. At a delay probability of $10^{-6}|V|$, \FroBarf{} is on average $2.0\times$, $2.6\times$, and $3.5\times$ faster on delay durations of $50$ ms, $100$ ms, and $200$ ms respectively. In \FroBarf{}, each thread dynamically selects a vertex chunk and updates their PageRank directly in a shared vector. If a thread is slow or faces random delays, another thread takes over to update remaining vertices in the next iteration without assigning new chunks to the delayed thread. This ensures eventual processing of all vertices. The accuracy of obtained PageRanks remains within an acceptable range compared to reference PageRanks, as shown in Figure \ref{fig:uniform-sleep-am-error}.

\subsection{Performance under Random Thread Crashes (Crash-stop model)}
\label{sec:fault-crash}

We also study how random thread crashes impact \FroBarf{}. We simulate non-uniform thread crashes by flagging a subset of threads as \texttt{crashed}\ignore{with a high probability of $10^{-5}$ per PageRank computation} at a random point in time during PageRank computation. We vary the number of threads that may crash as $0$, $1$, $2$, $4$, and from $8$ to $56$ in steps of $8$. On a graph with $10$ million vertices, this amounts to an average of $100$ crash instances per an iteration. Marking a thread as crashed means such a thread no longer participates in the computation. Such crashed threads do not introduce any malicious behavior and are equivalent to a thread experiencing an infinite delay (crash-stop model). We fix the batch size at $10^{-4} |E|$, note the time taken and the quality of PageRanks obtained by \FroBarf{} on the graphs listed in Table \ref{tab:dataset}. The results are shown if Figure \ref{fig:fault-crash}.

Due to iteration barriers, \FroWbar{} fails to complete the computation even if a single thread crashes. However, as Figures \ref{fig:nonuniform-crash-am-time} and \ref{fig:nonuniform-crash-am-error} show, \FroBarf{} experiences performance degradation with more crashed threads but still finishes computation, with almost no increase in error. The time required for PageRank computation is correlated with that illustrated in Figure \ref{fig:strong-scaling}, where the number of threads corresponds to the threads that did not crash. When $56$ out of $64$ threads crash, \FroBarf{} still operates at $40\%$ of its original speed. In the event of a thread crash, with \FroBarf{}, other threads update the ranks of unprocessed vertices in the next iteration. The completion time doesn't directly correlate with the ratio of maximum to available threads, as crashes are spread out during execution.

%% file: src/tab-dataset-temporal.tex
\begin{table}[hbtp]
  \centering
  \caption{List of 2 real-world dynamic graphs, i.e., temporal networks, obtained from the Stanford Large Network Dataset Collection \cite{snap14}. Here, $|V|$ is the number of vertices, $|E_T|$ the number of temporal edges (includes duplicate edges), and $|E|$ the number of static edges (with no duplicates).}
  \label{tab:dataset-temporal}
  \begin{tabular}{|c||c|c|c|c|}
    \toprule
    \textbf{Graph} &
    \textbf{\textbf{$|V|$}} &
    \textbf{\textbf{$|E_T|$}} &
    \textbf{\textbf{$|E|$}} \\
    \midrule
    wiki-talk-temporal & 1.14M & 7.83M & 3.31M \\ \hline
    sx-stackoverflow & 2.60M & 63.4M & 36.2M \\ \hline
  \bottomrule
  \end{tabular}
\end{table}

%% file: src/tab-dataset.tex
\begin{table}[hbtp]
  \centering
  \caption{List of $12$ graphs sourced from the SuiteSparse Matrix Collection \cite{suite19} ($*$ indicates directed graphs). Here, $|V|$ denotes the number of vertices, $|E|$ the number of edges (inclusive of self-loops), and $D_{avg}$ the average out-degree.}
  \label{tab:dataset}
  \begin{tabular}{|c||c|c|c|c|}
    \toprule
    \textbf{Graph} &
    \textbf{\textbf{$|V|$}} &
    \textbf{\textbf{$|E|$}} &
    \textbf{\textbf{$D_{avg}$}} \\
    \midrule
    \multicolumn{4}{|c|}{\textbf{Web Graphs (LAW)}} \\ \hline
    indochina-2004$^*$ & 7.41M & 199M & 26.8 \\ \hline  
    arabic-2005$^*$ & 22.7M & 654M & 28.8 \\ \hline  
    uk-2005$^*$ & 39.5M & 961M & 24.3 \\ \hline  
    webbase-2001$^*$ & 118M & 1.11B & 9.4 \\ \hline  
    it-2004$^*$ & 41.3M & 1.18B & 28.5 \\ \hline  
    sk-2005$^*$ & 50.6M & 1.98B & 39.1 \\ \hline  
    \multicolumn{4}{|c|}{\textbf{Social Networks (SNAP)}} \\ \hline
    com-LiveJournal & 4.00M & 73.4M & 18.3 \\ \hline  
    com-Orkut & 3.07M & 237M & 77.3 \\ \hline  
    \multicolumn{4}{|c|}{\textbf{Road Networks (DIMACS10)}} \\ \hline
    asia\_osm & 12.0M & 37.4M & 3.1 \\ \hline  
    europe\_osm & 50.9M & 159M & 3.1 \\ \hline  
    \multicolumn{4}{|c|}{\textbf{Protein k-mer Graphs (GenBank)}} \\ \hline
    kmer\_A2a & 171M & 531M & 3.1 \\ \hline  
    kmer\_V1r & 214M & 679M & 3.2 \\ \hline  
  \bottomrule
  \end{tabular}
\end{table}

%% file: src/fig-temporal.tex
\begin{figure}[!hbt]
  \centering
  \subfigure{
    \label{fig:temporal--am}
    \includegraphics[width=0.99\linewidth]{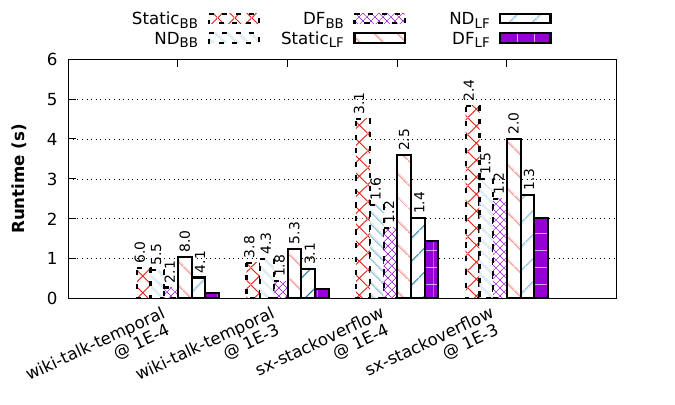}
  } \\[-2ex]
  \caption{Mean Runtime with \StaWbar{}, \NaiWbar{}, \FroWbar{}, \StaBarf{}, \NaiBarf{}, and \FroBarf{} on real-world dynamic graphs (see Table \ref{tab:dataset-temporal}), with batch updates of size $10^{-4}|E_T|$ and $10^{-3}|E_T|$. The labels on top of each bar indicates the speedup of \FroBarf{} over the respective approach.}
  \label{fig:temporal}
\end{figure}

%% file: src/fig-stability.tex
\begin{figure}[!hbt]
  \centering
  \subfigure{
    \label{fig:stability-am-key}
    \includegraphics[width=0.75\linewidth]{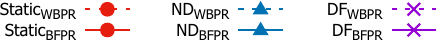}
  } \\[-1ex]
  \subfigure{
    \label{fig:stability-am}
    \includegraphics[width=0.65\linewidth]{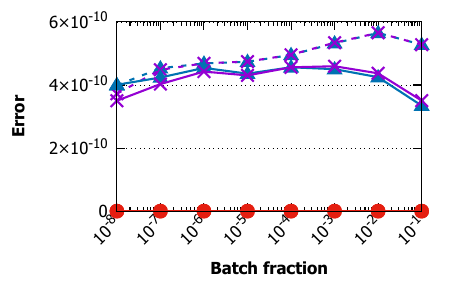}
  } \\[-2ex]
\caption{Error with \StaWbar{}, \NaiWbar{}, and \FroWbar{} (dashed lines) along with \StaBarf{}, \NaiBarf{}, and \FroBarf{} (solid lines) on deletions-only followed by insertions-only (which reverses the deletions) batch updates (X-axis) of increasing size for each graph in the dataset, with respect to \StaWbar{} and \StaBarf{} respectively on the original graph. The error is averaged over the graphs in Table \ref{tab:dataset}.}
  \label{fig:stability}
\end{figure}

%% file: src/fig-strong-scaling.tex
\begin{figure}[!hbt]
  \centering
  \subfigure{
    \label{fig:strong-scaling-all}
    \includegraphics[width=0.65\linewidth]{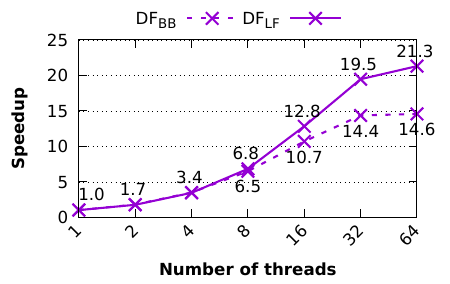}
  } \\[-2ex]
  \caption{Average speedup of \FroWbar{}\ignore{(dashed lines)} and \FroBarf{}\ignore{(solid lines)} with respect to sequential version of the same approach, over increasing thread count from $1$ to $64$ in multiples of $2$, on a fixed batch size of $10^{-4}|E|$ and in the absence of faults.}
  \label{fig:strong-scaling}
\end{figure}

%% file: src/fig-fault-none.tex
\begin{figure*}[!hbt]
  \centering
  \includegraphics[width=0.55\linewidth]{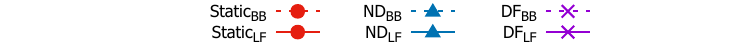} \\[-1ex]
  \begin{minipage}{0.66\linewidth}
  \subfigure[Results on each graph]{
    \label{fig:no-failure-all}
    \includegraphics[width=0.99\linewidth]{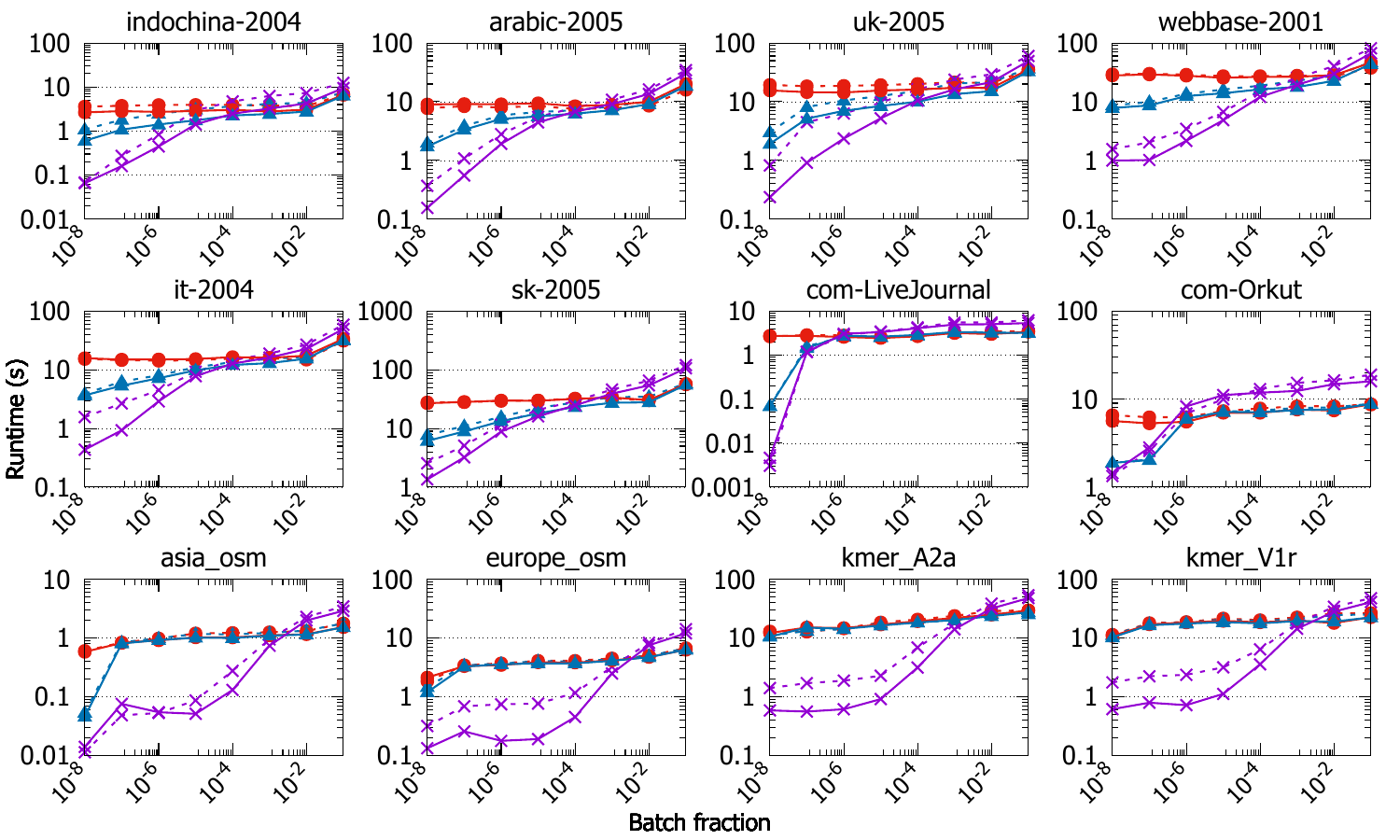}
  }
  \end{minipage}
  \begin{minipage}{0.32\linewidth}
    \subfigure[Overall runtime]{
      \label{fig:no-failure-am-time}
      \includegraphics[width=0.99\linewidth]{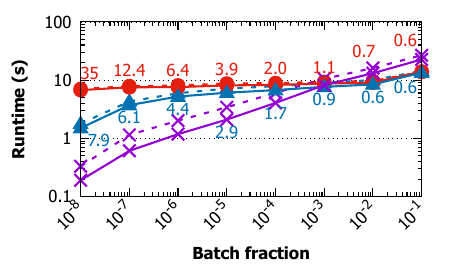}
    } \\[-2ex]
    \subfigure[Overall error]{
      \label{fig:no-failure-am-error}
      \includegraphics[width=0.99\linewidth]{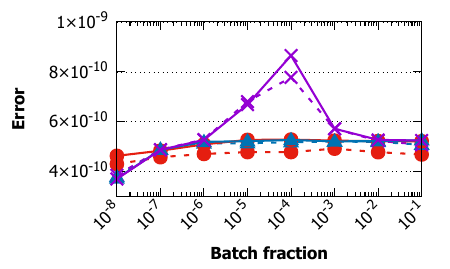}
    }
  \end{minipage} \\[-3ex]
  \caption{Runtime in seconds for \StaWbar{}, \NaiWbar{}, and \FroWbar{} (dashed lines\ignore{, Sections \ref{sec:withbarrier}, \ref{sec:naivedynamic-withbarrier}, and \ref{sec:frontier-withbarrier}}) along with \StaBarf{}, \NaiBarf{}, and \FroBarf{} (solid lines\ignore{, Sections \ref{sec:barrierfree}, \ref{sec:approach}, and \ref{sec:frontier-barrierfree}}) on batch updates of size $10^{-8}|E|$ to $0.1|E|$ (in multiples of $10$) for each graph, in the absence of faults. Mean runtime and error are shown on the right, with labels indicating speedup of \FroBarf{} with respect to \StaBarf{} and \NaiBarf{}.}
  \label{fig:no-failure}
\end{figure*}

%% file: src/fig-fault-sleep.tex
\begin{figure*}[!hbt]
  \centering
  \includegraphics[width=0.55\linewidth]{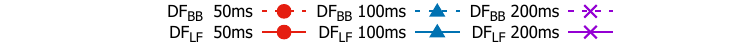} \\[-1ex]
  \begin{minipage}{0.66\linewidth}
  \subfigure[Results on each graph]{
    \label{fig:uniform-sleep-all}
    \includegraphics[width=0.99\linewidth]{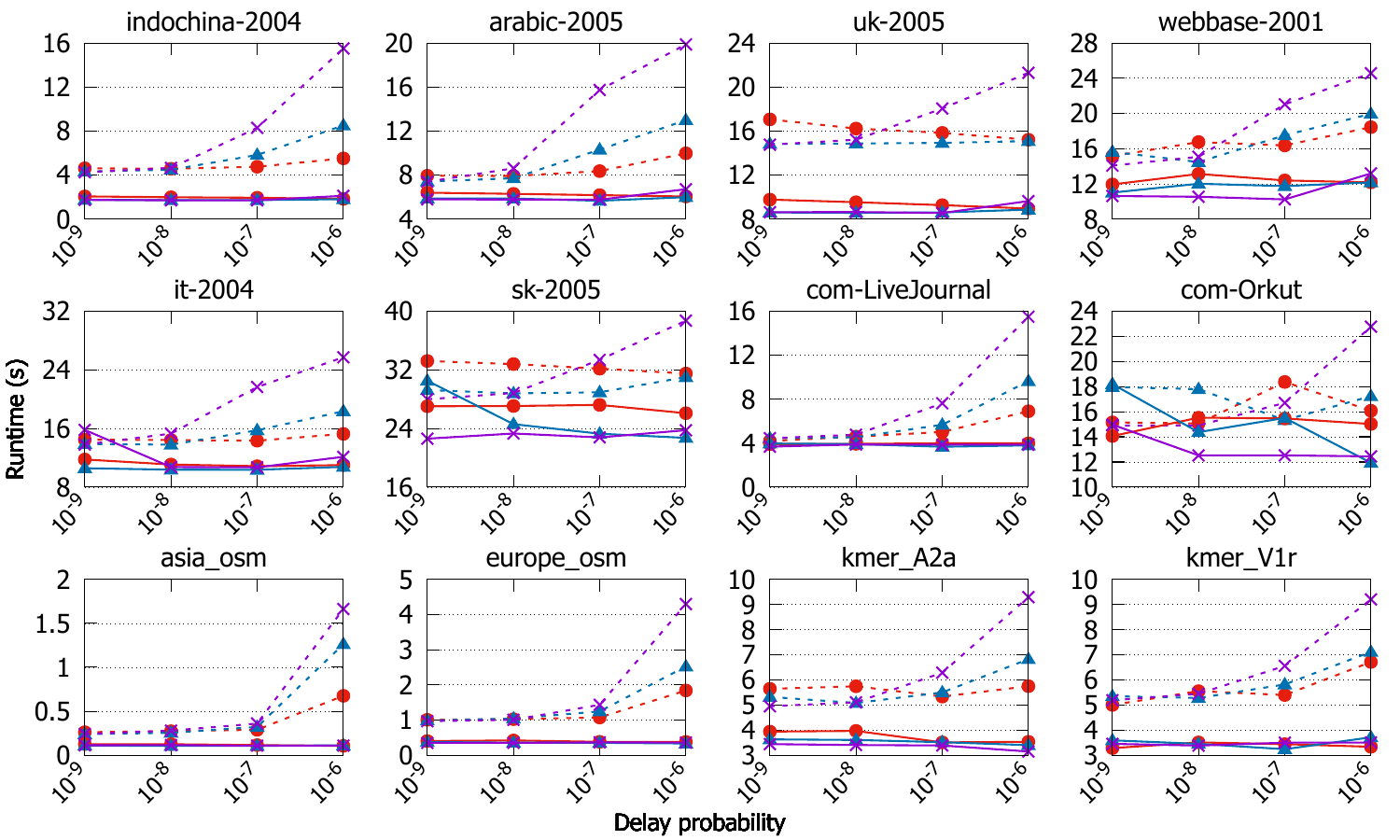}
  }
  \end{minipage}
  \begin{minipage}{0.32\linewidth}
  \subfigure[Overall runtime]{
    \label{fig:uniform-sleep-am-time}
    \includegraphics[width=0.99\linewidth]{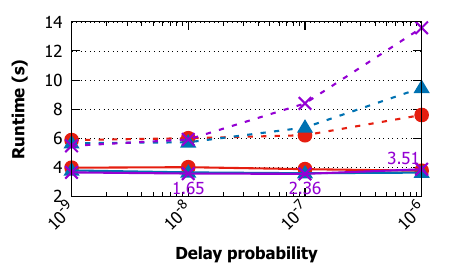}
  } \\[-2ex]
  \subfigure[Overall error]{
    \label{fig:uniform-sleep-am-error}
    \includegraphics[width=0.99\linewidth]{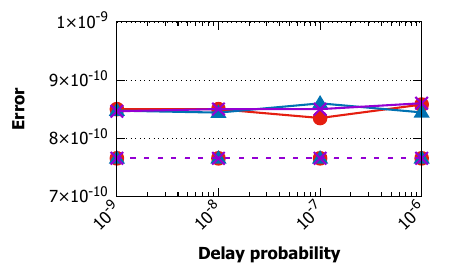}
  }
  \end{minipage} \\[-3ex]
  \caption{Runtime in seconds with \FroWbar{} (dashed lines) and \FroBarf{} (solid lines) on a batch size of $10^{-4} |E|$, and in the presence of random thread delays of $50$ ms, $100$ ms, and $200$ ms with probabilities of $10^{-9}|V|$ to $10^{-6}|V|$ (in multiples of $10$) per iteration. Mean runtime and error are shown on the right, with the labels indicating speedup of \FroBarf{} over \FroWbar{} in the presence of random thread delays of $50$ ms, $100$ ms, and $200$ ms.}
  \label{fig:fault-sleep}
\end{figure*}

%% file: src/fig-fault-crash.tex
\begin{figure*}[!hbt]
  \centering
  \includegraphics[width=0.55\linewidth]{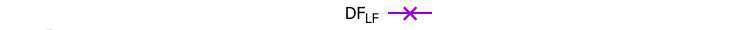} \\[-1ex]
  \begin{minipage}{0.66\linewidth}
  \subfigure[Results on each graph]{
    \label{fig:nonuniform-crash-all}
    \includegraphics[width=0.99\linewidth]{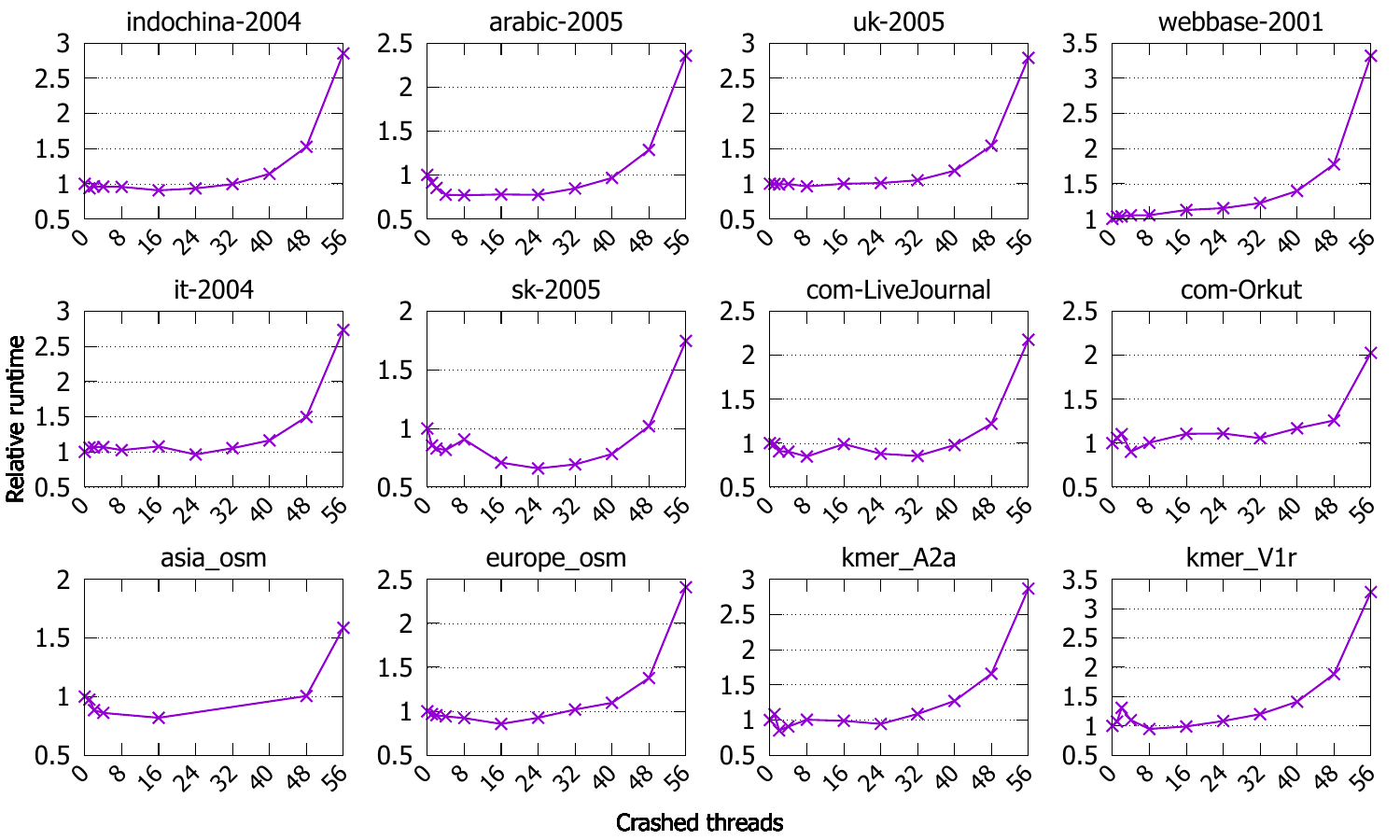}
  }
  \end{minipage}
  \begin{minipage}{0.32\linewidth}
  \subfigure[Overall relative runtime]{
    \label{fig:nonuniform-crash-am-time}
    \includegraphics[width=0.99\linewidth]{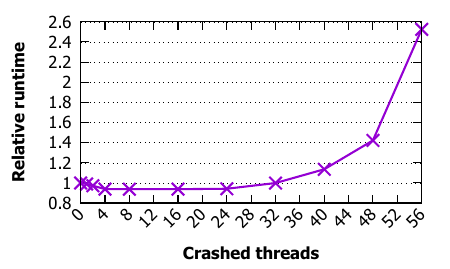}
  } \\[-2ex]
  \subfigure[Overall error]{
    \label{fig:nonuniform-crash-am-error}
    \includegraphics[width=0.99\linewidth]{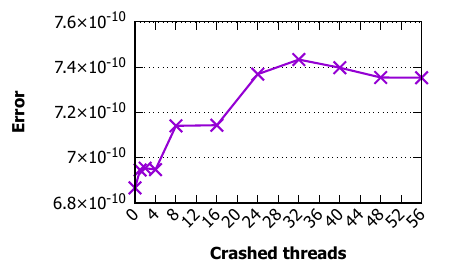}
  }
  \end{minipage} \\[-3ex]
  \caption{Relative runtime of \FroBarf{} on a batch size of $10^{-4} |E|$, and in the presence of random thread crashes\ignore{with a probability of $10^{-5}$ per rank computation}, with respect to runtime of \FroBarf{} with no crashed threads. Number of crashed threads is shown in X-axis with $0$, $1$, $2$, $4$, and from $8$ to $56$ in steps of $8$. Mean relative runtime and error are shown on the right.}
  \label{fig:fault-crash}
\end{figure*}

%% file: 06-conclusion.tex
In this report, we presented a lock-free PageRank algorithm for updating PageRank scores on dynamic graphs. First, we presented our Dynamic Frontier (DF) approach which incrementally identifies and processes vertices that are likely to change their PageRanks with minimal overhead. We integrated this with our lock-free PageRank (\FroBarf{}), with a helping mechanism among threads between its two phases. Experimental results show \FroBarf{} outperforms lock-free Naive-dynamic PageRank (\NaiBarf{}) by $2.5\times$ / $4.6\times$ on real-world dynamic graphs / large real-world graphs with random updates and is $1.6\times$ faster than barrier-based Dynamic Frontier PageRank (\FroWbar{}). While lock-free computations may introduce some redundancy, addressing it by reducing the chunk size impacts performance. Thus, a certain amount of redundancy in this case is tolerable.

\paragraph{Future Directions}

In this work we focus on dynamic graphs with edge insertions and deletions. For future research, we plan to extend the algorithm to handle vertex additions and deletions by scaling existing vertex ranks before computation.

%% file: aa-appendix.tex
\subsection{Barrier-based Static PageRank (\StaWbar{})}
\label{sec:describe-with-barrier-static}

The standard barrier-based Static PageRank, referred to as \StaWbar{}, is given in Algorithm \ref{alg:with-barrier-static}. It takes as input the current graph $G^t$ and returns the PageRanks $R$. The algorithm initializes the PageRank vectors $R$ and $R_{new}$, setting the initial PageRank of each vertex to $1/|V^t|$ (line \ref{alg:with-barrier-static--initialize}). Next, the algorithm iteratively computes the PageRank for each vertex $v$ (lines \ref{alg:with-barrier-static--iterations-begin}-\ref{alg:with-barrier-static--iterations-end}). Each iteration begins by computing the rank of vertex $v$ (lines \ref{alg:with-barrier-static--rank-begin}-\ref{alg:with-barrier-static--rank-end}). Next, the change in rank $\Delta r$ is calculated as the absolute difference between the new and old ranks, and the new rank is then assigned to $R_{new}[v]$ (line \ref{alg:with-barrier-static--update-rank}). Once all vertices have their ranks computed in parallel, an implicit barrier ensures all threads synchronize (line \ref{alg:with-barrier-static--rank-barrier}). Following this, the maximum change in ranks $\Delta R$ is computed using the $L_\infty$-norm between the previous iteration's ranks $R$ and the current iteration's ranks $R_{new}$ (line \ref{alg:with-barrier-static--rank-changes}). Another implicit barrier ensures synchronization after this computation. The vectors $R$ and $R_{new}$ are then swapped to set up the next iteration (line \ref{alg:with-barrier-static--swap}). The algorithm checks if the maximum change in ranks $\Delta R$ is less than or equal to the iteration tolerance $\tau$ (line \ref{alg:with-barrier-static--converged}). If so, the iterations stop as the ranks have converged. In the end, the PageRank vector $R$ is returned (line \ref{alg:with-barrier-static--return}).

\input{src/alg-with-barrier-static}
\input{src/alg-barrier-free-static}

\subsection{Lock-free Static PageRank (\StaBarf{})}
\label{sec:describe-barrier-free-static}

We now describe the implementation of our lock-free Static PageRank, which we refer to as \textit{Static}$_\text{LF}$, given in Algorithm \ref{alg:barrier-free-static}. It takes as input the current graph $G^t$ and returns the updated PageRanks $R$. In the algorithm, we begin by initializing the rank convergence vector $R_C$ to $0$ for all vertices, indicating that no vertex has yet converged, and the rank vector $R$ with the initial PageRank values (line \ref{alg:barrier-free-static--init}). Next, the algorithm proceeds with parallel execution of the iterative PageRank computation (lines \ref{alg:barrier-free-static--parallel-begin}-\ref{alg:barrier-free-static--iterations-end}). For a maximum of $MAX\_ITERATIONS$ iterations, each vertex $v \in V^t$ computes its new rank $r$ (lines \ref{alg:barrier-free-static--rank-begin}-\ref{alg:barrier-free-static--rank-end}). Next, the change in rank $\Delta r$ is calculated as the absolute difference between the new rank $r$ and the current rank $R[v]$, and the rank of vertex $v$ in the rank vector $R$ is then updated to the new rank $r$ (line \ref{alg:barrier-free-static--update}). If the change in rank $\Delta r$ is less than or equal to the iteration tolerance $\tau$, and the vertex $v$ was previously marked as not converged (i.e., $R_C[v] = 1$), the vertex $v$ is now marked as converged by setting $R_C[v]$ to $0$ (line \ref{alg:barrier-free-static--converged-check}). After updating the ranks for all vertices, the algorithm checks if the ranks have converged by verifying if all entries in $R_C$ are $0$ (line \ref{alg:barrier-free-static--converged}). If this condition is met, the iterations stop as the ranks have converged.\ignore{The iterations continue until either the maximum number of iterations $MAX\_ITERATIONS$ is reached or all ranks have converged.} Finally, the algorithm returns the final PageRank vector $R$ (line \ref{alg:barrier-free-static--return}).

\input{src/alg-with-barrier-naive-dynamic}

\subsection{Barrier-based ND PageRank (\NaiWbar{})}
\label{sec:describe-with-barrier-naive-dynamic}

We now describe the implementation of barrier-based Naive-dynamic PageRank, which we refer to as \NaiWbar{}, given in Algorithm \ref{alg:with-barrier-naive-dynamic}. It takes as input the current graph $G^t$ and the previous PageRank vector $R^{t-1}$, and returns the updated PageRanks $R^t = R$. In the algorithm, we begin by initializing the PageRank vectors $R$ and $R_{new}$ for consecutive iterations with the previous PageRank vector $R^{t-1}$ (line \ref{alg:with-barrier-naive-dynamic--init}). As mentioned earlier, we perform a synchronous PageRank computation with \NaiWbar{} using two PageRank vectors, $R$ and $R_{new}$. Next, we enter the main loop, iterating up to a maximum number of iterations $MAX\_ITERATIONS$ (lines \ref{alg:with-barrier-naive-dynamic--iterations-begin}-\ref{alg:with-barrier-naive-dynamic--iterations-end}). For each vertex $v$ in the current graph $G^t$, we compute the new PageRank $r$ in parallel (lines \ref{alg:with-barrier-naive-dynamic--rank-begin}-\ref{alg:with-barrier-naive-dynamic--rank-end}). Next, we calculate the change in rank $\Delta r$ as the absolute difference between the new rank $r$ and the current rank $R[v]$, and update $R_{new}[v]$ with the new rank $r$ (line \ref{alg:with-barrier-naive-dynamic--update}). Once all vertices have been processed, an implicit barrier ensures that all threads wait until the computation for the current iteration is complete (line \ref{alg:with-barrier-naive-dynamic--wait}). We then compute the maximum change in PageRanks $\Delta R$ between the previous iteration $R$ and the current iteration $R_{new}$ using the $L_\infty$-norm in parallel, with another implicit barrier (line \ref{alg:with-barrier-naive-dynamic--norm}). The vectors $R_{new}$ and $R$ are then swapped to set up for the next iteration (line \ref{alg:with-barrier-naive-dynamic--swap}). The iteration continues until the maximum change in PageRanks $\Delta R$ falls below the iteration tolerance $\tau$ (line \ref{alg:with-barrier-naive-dynamic--converged}), or the maximum number of iterations $MAX\_ITERATIONS$ is reached. Finally, the algorithm returns the final PageRank vector $R$\ignore{(line \ref{alg:with-barrier-naive-dynamic--return})}.

\input{src/alg-barrier-free-naive-dynamic}

\subsection{Our Lock-free ND PageRank (\NaiBarf{})}
\label{sec:describe-barrier-free-naive-dynamic}

We now describe our lock-free Naive-dynamic PageRank algorithm, which we refer to as \NaiBarf{}, given in Algorithm \ref{alg:barrier-free-naive-dynamic}. It takes as input the current graph $G^t$ and the previous PageRank vector $R^{t-1}$, and returns the updated PageRanks $R^t = R$. In the algorithm, we begin by initializing the convergence tracking vector $R_C$ to all zeros and setting the PageRank vector $R$ for the current iteration to the previous PageRank vector $R^{t-1}$. The algorithm runs the following steps in parallel. For each iteration $i$ (lines \ref{alg:barrier-free-naive-dynamic--iterations-begin}-\ref{alg:barrier-free-naive-dynamic--iterations-end}), we iterate over each vertex $v$ in the current graph $V^t$ using a dynamic scheduling approach (lines \ref{alg:barrier-free-naive-dynamic--vertex-begin}-\ref{alg:barrier-free-naive-dynamic--vertex-end}). The PageRank $r$ for each vertex $v$ is the computed (lines \ref{alg:barrier-free-naive-dynamic--rank-begin}-\ref{alg:barrier-free-naive-dynamic--rank-end}). After computing the new PageRank $r$ for vertex $v$, we calculate the change in rank $\Delta r$ as the absolute difference between the new and old PageRanks, and update the PageRank vector $R[v]$ with the new value $r$ (line \ref{alg:barrier-free-naive-dynamic--update}). If the change $\Delta r$ is less than or equal to the iteration tolerance $\tau$ and the rank of vertex $v$ has not yet converged (indicated by $R_C[v] = 1$), we mark vertex $v$ as converged by setting $R_C[v]$ to $0$ (line \ref{alg:barrier-free-naive-dynamic--converged-check}). At the end of each iteration, we check if all vertices have converged (i.e., $R_C[v] = 0$ for all $v$ in $V^t$). If so, the algorithm terminates early (line \ref{alg:barrier-free-naive-dynamic--converged}). Finally, the algorithm returns the updated PageRank vector $R$ (line \ref{alg:barrier-free-naive-dynamic--return}).

\input{src/alg-with-barrier-dynamic-traversal}

\subsection{Barrier-based DT PageRank (\TraWbar{})}
\label{sec:describe-with-barrier-dynamic-traversal}

We now describe our barrier-based Dynamic Traversal (DT) PageRank, which we refer to as $\text{DT}_{\text{BB}}$, given in Algorithm \ref{alg:with-barrier-dynamic-traversal}. It takes as input the previous graph $G^{t-1}$ and the current graph $G^t$, edge deletions $\Delta^{t-}$ and insertions $\Delta^{t+}$ in the batch update, the previous PageRank vector $R^{t-1}$, and returns the updated PageRanks $R^t = R$. In the algorithm, we begin by initializing the PageRank vectors $R$ and $R_{new}$ for consecutive iterations with the previous PageRank vector $R^{t-1}$, and initialize the affected vertices vector $V_A$ with zeros (line \ref{alg:with-barrier-dynamic-traversal--init}). Here, we perform a synchronous PageRank computation with $\text{DT}_{\text{BB}}$ using two PageRank vectors $R$ and $R_{new}$. Next, for each edge in the batch update, i.e., $\Delta^{t-}$ or $\Delta^{t+}$, we initially perform a depth-first search (DFS) traversal from all out-neighbors of the source vertex $u$ in both $G^{t-1}$ and $G^t$, marking the visited vertices as affected in parallel (lines \ref{alg:with-barrier-dynamic-traversal--affected-begin}-\ref{alg:with-barrier-dynamic-traversal--affected-begin}). This is done using the $visitDfs()$ function, which marks vertices reachable from a given starting vertex $v'$ in the current graph $G^t$. We then iteratively compute the PageRank $r$ for each affected vertex $v$ (lines \ref{alg:with-barrier-dynamic-traversal--vertex-begin}-\ref{alg:with-barrier-dynamic-traversal--vertex-end}). The PageRank $r$ for each vertex $v$ is the computed (lines \ref{alg:with-barrier-dynamic-traversal--rank-begin}-\ref{alg:with-barrier-dynamic-traversal--rank-end}). After computing the new PageRank $r$ for vertex $v$, we calculate the change in rank $\Delta r$ as the absolute difference between the new and old PageRanks, and update the PageRank vector $R[v]$ with the new value $r$ (line \ref{alg:with-barrier-dynamic-traversal--update}). There is an implicit iteration barrier at the end of the loop, where threads synchronize after computing PageRanks in parallel. Next, in parallel, we compute the maximum error $\Delta R$ between the PageRanks obtained in the previous $R$ and the current iteration $R_{new}$ using the $L_\infty$-norm. We then swap $R_{new}$ and $R$ to set up the next iteration. The iteration continues until either the maximum change in PageRanks $\Delta R$ falls below the iteration tolerance $\tau$\ignore{, or the maximum number of iterations $MAX\_ITERATIONS$ is reached}. At the end, the algorithm returns the final PageRank vector $R$ (line \ref{alg:with-barrier-dynamic-traversal--return}).

\input{src/alg-barrier-free-dynamic-traversal}

\subsection{Lock-free DT PageRank (\TraBarf{})}
\label{sec:describe-barrier-free-dynamic-traversal}

We now describe our lock-free Dynamic Traversal (DT) PageRank algorithm, referred to as \TraBarf{}, as given in Algorithm \ref{alg:barrier-free-dynamic-traversal}. This algorithm updates PageRanks dynamically and asynchronously without relying on barriers for synchronization. It takes as input the previous graph $G^{t-1}$, the previous PageRank vector $R^{t-1}$, and edge deletions $\Delta^{t-}$ and insertions $\Delta^{t+}$ in the batch update, and returns the updated PageRanks $R^t = R$. In the algorithm, we begin by initializing the vectors $C$, $V_A$, and $R_C$ to zero, which track whether vertices have been checked, are affected, or their ranks have not yet converged, respectively. We also initialize the PageRank vector $R$ with the previous PageRank vector $R^{t-1}$. Next, we mark the initially affected vertices due to the batch update in parallel, retrying on failure (lines \ref{alg:barrier-free-dynamic-traversal--affected-begin}-\ref{alg:barrier-free-dynamic-traversal--affected-end}). For each edge in $\Delta^{t-}$ or $\Delta^{t+}$, if the source vertex $u$ has not been marked as checked ($C[u] = 0$), we mark all out-neighbors of $u$ in both $G^{t-1}$ and $G^t$ as affected using a depth-first search via the function $visitDfs()$. After marking, we set $C[u] = 1$. This process continues until all vertices affected by the batch update have been marked. We then iteratively compute the PageRank $r$ for each vertex $v$ in the current graph $V^t$ in a dynamically scheduled manner (lines \ref{alg:barrier-free-dynamic-traversal--iterations-begin}-\ref{alg:barrier-free-dynamic-traversal--iterations-end}). If a vertex $v$ is not affected ($V_A[v] = 0$), we skip the computation for that vertex. For affected vertices, we compute their new PageRank considering the incoming edges $G^t.in(v)$. The change in PageRank $\Delta r$ is calculated, and the rank vector $R$ is updated. If the change $\Delta r$ is within the iteration tolerance $\tau$ and the vertex $v$ had not yet converged ($R_C[v] = 1$), we mark the vertex as converged by setting $R_C[v] = 0$. The iteration continues until all vertices' ranks have converged ($R_C[v] = 0$ for all $v \in V^t$)\ignore{or the maximum number of iterations $MAX\_ITERATIONS$ is reached}. Finally, the algorithm returns the updated PageRank vector $R$ (line \ref{alg:barrier-free-dynamic-traversal--return}).

%% file: src/alg-with-barrier-static.tex
\begin{algorithm}[!hbt]
\caption{Barrier-based Static PageRank\ignore{($\text{Static}_{\text{BB}}$)}.}
\label{alg:with-barrier-static}
\begin{algorithmic}[1]
\Require{$G^t$: Input graph}
\Ensure{$R, R_{new}$: Rank vectors in current, previous iteration}
\Ensure{$\Delta R$: $L_\infty$-norm between ranks in consecutive iterations}
\Ensure{$\Delta r$: Change in rank of a vertex}
\Ensure{$\tau$: Iteration tolerance}
\Ensure{$\alpha$: Damping factor}

\Statex

\Function{$\text{Static}_{\text{BB}}$}{$G^t$}
  \State $R_{new} \gets R \gets \{1/|V^t| \forall\ v \in V^t\}$ \label{alg:with-barrier-static--initialize}
  \ForAll{$i \in [0 .. MAX\_ITERATIONS)$} \label{alg:with-barrier-static--iterations-begin}
    \ForAll{$v \in V^t$ \textbf{in dynamic parallel}}
      \State $r \gets (1 - \alpha)/|V^t|$ \label{alg:with-barrier-static--rank-begin}
      \ForAll{$u \in G^t.in(v)$}
        \State $r \gets r + \alpha * R[u] / |G^t.out(u)|$
      \EndFor \label{alg:with-barrier-static--rank-end}
      \State $\Delta r \gets |r - R[v]|$ \textbf{;} $R_{new}[v] \gets r$ \label{alg:with-barrier-static--update-rank}
    \EndFor
    \State \textbf{wait for all threads} (implicit barrier) \label{alg:with-barrier-static--rank-barrier}
    \State $\Delta R \gets l_\infty Norm(R, R_{new})$ \textbf{in parallel} (implicit barrier)\label{alg:with-barrier-static--rank-changes}
    \State $swap(R_{new}, R)$ \label{alg:with-barrier-static--swap}
    \State $//$ Ranks converged?
    \If{$\Delta R \le \tau$} \textbf{break} \label{alg:with-barrier-static--converged}
    \EndIf
  \EndFor \label{alg:with-barrier-static--iterations-end}
  \Return{$R$} \label{alg:with-barrier-static--return}
\EndFunction
\end{algorithmic}
\end{algorithm}

%% file: src/alg-barrier-free-static.tex
\begin{algorithm}[!hbt]
\caption{Lock-free Static PageRank\ignore{($\text{Static}_{\text{LF}}$)}.}
\label{alg:barrier-free-static}
\begin{algorithmic}[1]
\Require{$G^t$: Input graph}
\Ensure{$R$: Rank vector in the current iteration}
\Ensure{$R_C[u]$: Has the rank of vertex $u$ not yet converged?}
\Ensure{$\Delta r$: Change in rank of a vertex}
\Ensure{$\tau$: Iteration tolerance}
\Ensure{$\alpha$: Damping factor}

\Statex

\Function{$\text{Static}_{\text{LF}}$}{$G^t$}
  \State $R_C \gets \{0\}$ \textbf{;} $R \gets \{1/|V^t|\}$ \label{alg:barrier-free-static--init}
  \State \textbf{run steps below in parallel} \label{alg:barrier-free-static--parallel-begin}
    \ForAll{$i \in [0 .. MAX\_ITERATIONS)$} \label{alg:barrier-free-static--iterations-begin}
      \ForAll{$v \in V^t$ \textbf{in dynamic schedule}}
        \State $r \gets (1 - \alpha)/|V^t|$ \label{alg:barrier-free-static--rank-begin}
        \ForAll{$u \in G^t.in(v)$}
          \State $r \gets r + \alpha * R[u] / |G^t.out(u)|$
        \EndFor \label{alg:barrier-free-static--rank-end}
        \State $\Delta r \gets |r - R[v]|$ \textbf{;} $R[v] \gets r$ \label{alg:barrier-free-static--update}
        \If{$\Delta r \le \tau$ \textbf{and} $R_C[v] = 1$} $R_C[v] \gets 0$ \label{alg:barrier-free-static--converged-check}
        \EndIf
      \EndFor
      \State $//$ Ranks converged?
      \If{$R_C[v] = 0\ \forall\ v \in V^t$} \textbf{break} \label{alg:barrier-free-static--converged}
      \EndIf
    \EndFor \label{alg:barrier-free-static--iterations-end}
  \Return{$R$} \label{alg:barrier-free-static--return}
\EndFunction
\end{algorithmic}
\end{algorithm}

%% file: src/alg-with-barrier-naive-dynamic.tex
\begin{algorithm}[!hbt]
\caption{Barrier-based Naive-dynamic PageRank\ignore{($\text{ND}_{\text{BB}}$)}.}
\label{alg:with-barrier-naive-dynamic}
\begin{algorithmic}[1]
\Require{$G^t$: Current input graph}
\Require{$R^{t-1}$: Previous rank vector}
\Ensure{$R, R_{new}$: Rank vectors in current, previous iteration}
\Ensure{$\Delta R$: $L_\infty$-norm between ranks in consecutive iterations}
\Ensure{$\Delta r$: Change in rank of a vertex}
\Ensure{$\tau$: Iteration tolerance}
\Ensure{$\alpha$: Damping factor}

\Statex

\Function{$\text{ND}_{\text{BB}}$}{$G^t, R^{t-1}$}
  \State $R_{new} \gets R \gets R^{t-1}$ \label{alg:with-barrier-naive-dynamic--init}
  \ForAll{$i \in [0 .. MAX\_ITERATIONS)$} \label{alg:with-barrier-naive-dynamic--iterations-begin}
    \ForAll{$v \in V^t$ \textbf{in dynamic parallel}}
      \State $r \gets (1 - \alpha)/|V^t|$ \label{alg:with-barrier-naive-dynamic--rank-begin}
      \ForAll{$u \in G^t.in(v)$}
        \State $r \gets r + \alpha * R[u] / |G^t.out(u)|$
      \EndFor \label{alg:with-barrier-naive-dynamic--rank-end}
      \State $\Delta r \gets |r - R[v]|$ \textbf{;} $R_{new}[v] \gets r$ \label{alg:with-barrier-naive-dynamic--update}
    \EndFor
    \State \textbf{wait for all threads} (implicit barrier) \label{alg:with-barrier-naive-dynamic--wait}
    \State $\Delta R \gets l_\infty Norm(R, R_{new})$ \textbf{in parallel} (implicit barrier)\label{alg:with-barrier-naive-dynamic--norm}
    \State $swap(R_{new}, R)$ \label{alg:with-barrier-naive-dynamic--swap}
    \State $//$ Ranks converged?
    \If{$\Delta R \le \tau$} \textbf{break} \label{alg:with-barrier-naive-dynamic--converged}
    \EndIf
  \EndFor \label{alg:with-barrier-naive-dynamic--iterations-end}
  \Return{$R$} \label{alg:with-barrier-naive-dynamic--return}
\EndFunction
\end{algorithmic}
\end{algorithm}

%% file: src/alg-barrier-free-naive-dynamic.tex
\begin{algorithm}[!hbt]
\caption{Our Lock-free Naive-dynamic PageRank\ignore{($\text{ND}_{\text{LF}}$)}.}
\label{alg:barrier-free-naive-dynamic}
\begin{algorithmic}[1]
\Require{$G^t$: Current input graph}
\Require{$R^{t-1}$: Previous rank vector}
\Ensure{$R$: Rank vector in the current iteration}
\Ensure{$R_C[u]$: Has the rank of vertex $u$ not yet converged?}
\Ensure{$\Delta r$: Change in rank of a vertex}
\Ensure{$\tau$: Iteration tolerance}
\Ensure{$\alpha$: Damping factor}

\Statex

\Function{$\text{ND}_{\text{LF}}$}{$G^t, R^{t-1}$}
  \State $R_C \gets \{0\}$ \textbf{;} $R \gets R^{t-1}$
  \State \textbf{run steps below in parallel}
    \ForAll{$i \in [0 .. MAX\_ITERATIONS)$} \label{alg:barrier-free-naive-dynamic--iterations-begin}
      \ForAll{$v \in V^t$ \textbf{in dynamic schedule}} \label{alg:barrier-free-naive-dynamic--vertex-begin}
        \State $r \gets (1 - \alpha)/|V^t|$ \label{alg:barrier-free-naive-dynamic--rank-begin}
        \ForAll{$u \in G^t.in(v)$}
          \State $r \gets r + \alpha * R[u] / |G^t.out(u)|$
        \EndFor \label{alg:barrier-free-naive-dynamic--rank-end}
        \State $\Delta r \gets |r - R[v]|$ \textbf{;} $R[v] \gets r$ \label{alg:barrier-free-naive-dynamic--update}
        \If{$\Delta r \le \tau$ \textbf{and} $R_C[v] = 1$} $R_C[v] \gets 0$ \label{alg:barrier-free-naive-dynamic--converged-check}
        \EndIf
      \EndFor \label{alg:barrier-free-naive-dynamic--vertex-end}
      \State $//$ Ranks converged?
      \If{$R_C[v] = 0\ \forall\ v \in V^t$} \textbf{break} \label{alg:barrier-free-naive-dynamic--converged}
      \EndIf
    \EndFor \label{alg:barrier-free-naive-dynamic--iterations-end}
  \Return{$R$} \label{alg:barrier-free-naive-dynamic--return}
\EndFunction
\end{algorithmic}
\end{algorithm}

%% file: src/alg-with-barrier-dynamic-traversal.tex
\begin{algorithm}[!hbt]
\caption{Barrier-based Dynamic Traversal PageRank\ignore{($\text{DT}_{\text{BB}}$)}.}
\label{alg:with-barrier-dynamic-traversal}
\begin{algorithmic}[1]
\Require{$G^{t-1}, G^t$: Previous, current input graph}
\Require{$R^{t-1}$: Previous rank vector}
\Require{$\Delta^{t-}, \Delta^{t+}$: Edge deletions and insertions}
\Ensure{$V_A[u]$: Is vertex $u$ affected due to current batch update?}
\Ensure{$R, R_{new}$: Rank vectors in current, previous iteration}
\Ensure{$\Delta R$: $L_\infty$-norm between ranks in consecutive iterations}
\Ensure{$\Delta r$: Change in rank of a vertex}
\Ensure{$\tau$: Iteration tolerance}
\Ensure{$\alpha$: Damping factor}

\Statex

\Function{$\text{DT}_{\text{BB}}$}{$G^{t-1}, R^{t-1}, \Delta^{t-}, \Delta^{t+}$}
  \State $V_A \gets \{0\}$ \textbf{;} $R_{new} \gets R \gets R^{t-1}$ \label{alg:with-barrier-dynamic-traversal--init}
  \State $//$ Mark initial affected
  \ForAll{$(u, v) \in \Delta^{t-} \cup \Delta^{t+} \textbf{in parallel}$} \label{alg:with-barrier-dynamic-traversal--affected-begin}
    \ForAll{$v' \in G^{t-1}.out(u) \cup G^t.out(u)$}
      \State $visitDfs(V_A, G^t, v')$
    \EndFor
  \EndFor \label{alg:with-barrier-dynamic-traversal--affected-end}
  \State \textbf{wait for all threads} (implicit barrier)
  \ForAll{$i \in [0 .. MAX\_ITERATIONS)$}
    \ForAll{$v \in V^t$ \textbf{in dynamic parallel}} \label{alg:with-barrier-dynamic-traversal--vertex-begin}
      \State $//$ Is vertex not affected?
      \If{$V_A[v] = 0$} \textbf{continue}
      \EndIf
      \State $r \gets (1 - \alpha)/|V^t|$ \label{alg:with-barrier-dynamic-traversal--rank-begin}
      \ForAll{$u \in G^t.in(v)$}
        \State $r \gets r + \alpha * R[u] / |G^t.out(u)|$
      \EndFor \label{alg:with-barrier-dynamic-traversal--rank-end}
      \State $\Delta r \gets |r - R[v]|$ \textbf{;} $R_{new}[v] \gets r$ \label{alg:with-barrier-dynamic-traversal--update}
    \EndFor \label{alg:with-barrier-dynamic-traversal--vertex-end}
    \State \textbf{wait for all threads} (implicit barrier)
    \State $\Delta R \gets l_\infty Norm(R, R_{new})$ \textbf{in parallel}
    \State $swap(R_{new}, R)$
    \State $//$ Ranks converged?
    \If{$\Delta R \le \tau$} \textbf{break}
    \EndIf
  \EndFor
  \Return{$R$} \label{alg:with-barrier-dynamic-traversal--return}
\EndFunction
\end{algorithmic}
\end{algorithm}

%% file: src/alg-barrier-free-dynamic-traversal.tex
\begin{algorithm}[!hbt]
\caption{Lock-free Dynamic Traversal PageRank\ignore{($\text{DT}_{\text{BB}}$)}.}
\label{alg:barrier-free-dynamic-traversal}
\begin{algorithmic}[1]
\Require{$G^{t-1}, G^t$: Previous, current input graph}
\Require{$R^{t-1}$: Previous rank vector}
\Require{$\Delta^{t-}, \Delta^{t+}$: Edge deletions and insertions}
\Ensure{$V_A[u]$: Is vertex $u$ affected due to current batch update?}
\Ensure{$R_C[u]$: Has the rank of vertex $u$ not yet converged?}
\Ensure{$C[u]$: Has vertex $u$ from batch update been checked?}
\Ensure{$R, R_{new}$: Rank vectors in current, previous iteration}
\Ensure{$\Delta R$: $L_\infty$-norm between ranks in consecutive iterations}
\Ensure{$\Delta r$: Change in rank of a vertex}
\Ensure{$\tau$: Iteration tolerance}
\Ensure{$\alpha$: Damping factor}

\Statex

\Function{$\text{DT}_{\text{BB}}$}{$G^{t-1}, R^{t-1}, \Delta^{t-}, \Delta^{t+}$}
  \State $C \gets V_A \gets R_C \gets \{0\}$ \textbf{;} $R \gets R^{t-1}$
  \State \textbf{run steps below in parallel}
    \State $//$ Mark initial affected (retry on failure)
    \While{$true$} \label{alg:barrier-free-dynamic-traversal--affected-begin}
      \ForAll{$(u, v) \in \Delta^{t-} \cup \Delta^{t+} \textbf{in parallel}$}
        \State $//$ Not marked yet?
        \If{$C[u] = 0$}
          \ForAll{$v' \in G^{t-1}.out(u) \cup G^t.out(u)$}
            \State $visitDfs(V_A\ \&\ R_C, G^t, v')$
          \EndFor
        \EndIf
        \State $C[u] \gets 1$
      \EndFor
      \State $//$ All marked?
      \If{$C[u] = 1\ \forall\ (u, v) \in \Delta^{t-} \cup \Delta^{t+}$} \textbf{break}
      \EndIf
    \EndWhile \label{alg:barrier-free-dynamic-traversal--affected-end}
    \ForAll{$i \in [0 .. MAX\_ITERATIONS)$} \label{alg:barrier-free-dynamic-traversal--iterations-begin}
      \ForAll{$v \in V^t$ \textbf{in dynamic schedule}}
        \State $//$ Is vertex not affected?
        \If{$V_A[v] = 0$} \textbf{continue}
        \EndIf
        \State $r \gets (1 - \alpha)/|V^t|$
        \ForAll{$u \in G^t.in(v)$}
          \State $r \gets r + \alpha * R[u] / |G^t.out(u)|$
        \EndFor
        \State $\Delta r \gets |r - R[v]|$ \textbf{;} $R[v] \gets r$
        \If{$\Delta r \le \tau$ \textbf{and} $R_C[v] = 1$} $R_C[v] \gets 0$
        \EndIf
      \EndFor
      \State $//$ Ranks converged?
      \If{$R_C[v] = 0\ \forall\ v \in V^t$} \textbf{break}
      \EndIf
    \EndFor \label{alg:barrier-free-dynamic-traversal--iterations-end}
  \Return{$R$} \label{alg:barrier-free-dynamic-traversal--return}
\EndFunction
\end{algorithmic}
\end{algorithm}